\documentclass[letterpaper,preprintnumbers,twocolumn,eqsecnum,superscriptaddress,aps,nofootinbib]{revtex4}
\usepackage{amssymb}\usepackage[centertags]{amsmath}\usepackage{txfonts}\usepackage{epsfig}\usepackage{bm}
\usepackage{color}\usepackage{graphicx,graphics}\usepackage{multirow}\usepackage{float}\usepackage{ulem}
\usepackage[dvipdfm]{hyperref}\usepackage{setspace}\usepackage{slashed}\usepackage{booktabs}
\hypersetup{colorlinks=true, citecolor=blue, linkcolor=red, filecolor=black,urlcolor=blue}
\allowdisplaybreaks[4]

\begin{document}

\title{Parity violating  semi-inclusive deeply inelastic scattering at the Electron-Ion Collider}
\author{Kai-bao Chen}
\affiliation{School of Science, Shandong Jianzhu University, Jinan, Shandong 250101, China}
\author{Wei-hua Yang}\thanks{Corresponding author}
\affiliation{College of Nuclear Equipment and Nuclear Engineering, Yantai University, Yantai, Shandong 264005, China}

\begin{abstract}
We present a systematic calculation of the current jet production semi-inclusive deeply inelastic scattering process at the Electron-Ion Collider energy.
Contributions from weak interactions are considered which give rise to parity violating effects.
We consider the general form of the polarized electron beam scattering off the polarized target which has spin 1.
The calculations are carried out up to twist-3 level in the quantum chromodynamics parton model by applying the collinear expansion
where multiple gluon scattering is taken into account and gauge links are obtained automatically.
We present complete results for structure functions and spin/azimuthal asymmetries in terms of the gauge invariant transverse momentum dependent parton distribution functions.
Both the unpolarized and the polarized electron beam cases correspond to 24 azimuthal asymmetries,
in which 6 of them appear at the leading twist, while 18 of them contribute at twist-3 level.
In addition we also calculate the parity-violating asymmetries which arise from the interference of the electromagnetic and weak interactions.
\end{abstract}

\maketitle

\section{Introduction}

Thanks to the asymptotic freedom of the strong interaction, many high energy reactions can be studied within the formalism of quantum chromodynamics (QCD) factorization theorems \cite{Collins:1989gx}, which separate the calculable hard parts from the non-perturbative soft parts in the cross sections.
These soft parts often involve  parton distribution functions (PDFs) and fragmentation functions (FFs).
Both of them are important quantities in describing high energy reactions.
When three dimensional, i.e., the transverse momentum dependent (TMD) PDFs and FFs are considered, the sensitive quantities studied in experiments are often different azimuthal asymmetries.
These asymmetries are measurable quantities which can be used to extract TMD PDFs and FFs which give information about the nucleon structure and the hadronization mechanism.
If only TMD PDFs are taken into account, one of the best reactions to study them is the semi-inclusive deeply inelastic scattering (SIDIS) with current jet production process.
Usually one photon approximation is used to calculate the (SI)DIS processes.
When weak interaction is taken into account, considering the neutral current interactions, the intermediate propagator can either be virtual photon ($\gamma^*$) or $Z^0$ boson.
Since weak interaction does not respect parity conservation, we can study the asymmetries induced by the parity-violating effects through weak interaction.
This process is called as parity-violating deeply inelastic scattering (PVDIS).
Parity-violating asymmetries~\cite{Cahn:1977uu,Anselmino:1993tc}, arising from the interference of electromagnetic (EM) and weak interactions, were first observed in DIS experiments 
carried out at SLAC~\cite{Prescott:1978tm,Prescott:1979dh} and have been studied widely.
Recently, measurements have been carried out~\cite{Aniol:2004hp,Aniol:2005zf,Aniol:2005zg,Armstrong:2005hs,Androic:2009aa, Wang:2013kkc,Wang:2014guo,Anthony:2003ub,Anthony:2005pm,Spayde:1999qg,Ito:2003mr,Maas:2004dh,Maas:2004ta}.
Proposals for precise measurements in the future are available~\cite{PVDIS:Jlab6,PVDIS:JLab12}.
We further extend the consideration into the SIDIS at the Electron-Ion Collider (EIC)~\cite{Accardi:2012qut} energies in this paper.

The EIC is a high-energy, high-luminosity collider with the capability to accelerate polarized electron and nucleon/ions.
The high energy and luminosity combined with polarized beams will provide a wealth of data in an area never explored before.
Therefore, it offers many opportunities to study spin effects and different azimuthal asymmetries.
Though, the EIC is proposed mainly for the study of strong interactions, it has the ability to measure parity violating quantities when weak interaction is taken into account.
Electro-weak inclusive and semi-inclusive DIS processes have been studied extensively \cite{Anselmino:1993tc,Ji:1993ey,Anselmino:1994gn,Boer:1999uu,Anselmino:2001ey,deFlorian:2012wk,Moreno:2014kia}.
However, systematic researches about structure functions and spin/azimuthal asymmetries are still lacking.
This includes a full kinematic analysis for the cross section, QCD parton model calculations beyond the leading power accuracy and the study of hadron polarization effects, etc.

Higher twist effects are often significant for semi-inclusive reaction processes and TMD observables.
Especially for the case of twist-3 (sub-leading power) corrections, they often lead to azimuthal asymmetries which are different from the leading twist ones~\cite{Mulders:1995dh,Bacchetta:2006tn,Boer:1997mf}.
Thus, the studies of higher twist effects will give complementary or even direct access to the nucleon structure or hadronization mechanism.
It has been shown that the collinear expansion is a powerful tool to calculate higher twist effects systematically by taking into account multiple gluon exchange contributions.
By using collinear expansion, on the one hand gauge links will be generated automatically which make the calculation explicitly gauge invariant.
On the other hand the formalism takes a very simple factorization form which consists of calculable hard parts and TMD PDFs/FFs.
This will greatly simplify the systematic calculation of higher twist contributions.
Based on the collinear expansion formalism, higher twist contributions to DIS and electron positron annihilation processes have been studied extensively~\cite{Ellis:1982wd,Qiu:1990xxa,Liang:2006wp,Song:2013sja,Song:2010pf,Wei:2013csa,Wei:2014pma,Chen:2016moq,Wei:2016far,Yang:2017sxz,Yang:2020ezt}.

The rest of this paper is organized as follows. In Sec.~\ref{sec:formalism}, we make kinematic analysis for $ e^-N \to e^-q(\rm jet) X$ process and present the differential cross section in terms of structure functions.
In Sec.~\ref{sec:partonmodel} and \ref{sec:crosssection}, we present detailed calculations of the hadronic tensor and the cross section respectively up to twist-3 level in terms of the gauge invariant TMD PDFs in the QCD parton model.
The results including structure functions and spin/azimuthal asymmetries are given in Sec.~\ref{sec:result}.
Finally, a summary is given in Sec.~\ref{sec:summary}.

\section{The process and general form of the cross section} \label{sec:formalism}

\subsection{The semi-inclusive PVDIS process}

To be explicit, we consider the current jet production SIDIS process at EIC energies,
\begin{align}
e^-(l,\lambda_e) + N(p,S) \rightarrow e^-(l^\prime) + q(k^\prime) + X,
\end{align}
where $N$ can be a nucleon with spin-1/2 or an ion, e.g., a deuteron with spin-1.
$q$ denotes a quark which
corresponds to a jet of hadrons observed in experiments.
In this paper, we consider the case of the electron scattering off a spin-1 target.
This gives us the opportunity to access also the tensor polarization effects.
We consider the neutral current interaction at the tree level of electroweak theory, i.e., the exchange of a virtual photon $\gamma^*$ or a $Z^0$ boson with momentum $q=l-l'$ between the electron and the target.
The standard variables for SIDIS are
\begin{align}
Q^2 = -q^2,\  x_B=\frac{Q^2}{2 p\cdot q}, \  y=\frac{p\cdot q}{p \cdot l},\  s=(p+l)^2.
\label{eq:SIDIS-var}
\end{align}
The differential cross-section is given by
\begin{align}
  d\sigma = \frac{\alpha_{\rm em}^2}{sQ^4}A_r L^r_{\mu\nu}(l,\lambda_e, l^\prime)W_r^{\mu\nu}(q,p,S,k^\prime)\frac{d^3 l^\prime d^3 k^\prime}{(2\pi)^32E_{l^\prime} E_{k^\prime}}. \label{f:crosssec}
\end{align}
The symbol $r$ can be $\gamma\gamma$, $ZZ$ and $\gamma Z$, for EM, weak and interference terms, respectively.
A summation over $r$ in Eq. (\ref{f:crosssec}) is understood, i.e., the total cross section is given by
\begin{align}
  d\sigma = d\sigma^{ZZ} + d\sigma^{\gamma Z} + d\sigma^{\gamma\gamma}.
\end{align}
$A_r$'s are defined as
\begin{align}
& A_{\gamma\gamma} = e_q^2, \nonumber\\
& A_{ZZ} = \frac{Q^4}{\left[(Q^2+M_Z^2)^2 + \Gamma_Z^2 M_Z^2 \right] \sin^4 2\theta_W} \equiv \chi, \nonumber\\
& A_{\gamma Z} = \frac{2e_q Q^2 (Q^2+M_Z^2)}{\left[(Q^2+M_Z^2)^2 + \Gamma_Z^2 M_Z^2 \right] \sin^2 2\theta_W}\equiv \chi_{int}.
\end{align}
The leptonic tensors are given by
\begin{align}
 &L^{\gamma\gamma}_{\mu\nu}(l,\lambda_e, l^\prime)= 2\left[ l_\mu l^\prime_\nu + l_\nu l^\prime_\mu - (l\cdot l^\prime)g_{\mu\nu}  \right] + 2i\lambda_e \varepsilon_{\mu\nu l l^\prime}, \\
 & L^{ZZ}_{\mu\nu}(l,\lambda_e, l^\prime) =(c_1^e - c_3^e \lambda_e)L^{\gamma\gamma}_{\mu\nu}(l,\lambda_e, l^\prime), \\
 & L^{\gamma Z}_{\mu\nu}(l,\lambda_e, l^\prime)=(c_V^e - c_A^e \lambda_e) L^{\gamma\gamma}_{\mu\nu}(l,\lambda_e, l^\prime),
\end{align}
where $c_1^e = (c_V^e)^2 + (c_A^e)^2$ and $c_3^e = 2 c_V^e c_A^e$.
$c_V^e$ and $c_A^e$ are defined in the weak interaction current
$J_\mu (x)=\bar \psi(x)\Gamma_\mu\psi(x)$ with $\Gamma_\mu= \gamma_\mu (c_V^e - c_A^e \gamma^5)$.
Similar notations are also used for quarks where the superscript $e$ is replaced by $q$.
The hadronic tensors are given by
\begin{align}
  W_{\gamma\gamma}^{\mu\nu}&(q,p,S,k^\prime) = \sum_X (2\pi)^3 \delta^4(p + q - k^\prime - p_X)\nonumber\\
 &\quad \times\langle p,S |J_{\gamma\gamma}^\mu(0)|k^\prime;X\rangle \langle k^\prime;X |J_{\gamma\gamma}^\nu(0)| p,S \rangle , \\
  W_{ZZ}^{\mu\nu}&(q,p,S,k^\prime) = \sum_X (2\pi)^3 \delta^4(p + q - k^\prime - p_X)\nonumber\\
 &\quad \times\langle p,S | J_{ZZ}^\mu(0)|k^\prime;X\rangle \langle k^\prime;X | J_{ZZ}^\nu(0) | p,S \rangle , \\
  W_{\gamma Z}^{\mu\nu}&(q,p,S,k^\prime) = \sum_X (2\pi)^3 \delta^4(p + q - k^\prime - p_X)\nonumber\\
 &\quad \times\langle p,S |J_{ZZ}^\mu(0)|k^\prime;X\rangle \langle k^\prime;X | J_{\gamma\gamma}^\nu(0)| p,S \rangle ,
\end{align}
where $J_{\gamma\gamma}^\mu(0) =\bar\psi(0) \gamma^\mu \psi(0)$, $ J_{ZZ}^\mu(0) =\bar\psi(0) \Gamma^\mu_q \psi(0)$ with $\Gamma^\mu_q = \gamma^\mu(c_V^q - c_A^q \gamma_5)$.
$W_{r}^{\mu\nu}(q,p,S,k^\prime)$ is related to the hadronic tensor $W_{r}^{\mu\nu(in)}(q,p,S)$ for the inclusive process $e^- N \rightarrow e^- X$ by
\begin{align}
W_{r}^{\mu\nu(in)}(q,p,S) =\int \frac{d^3 k^\prime}{(2\pi)^32E_{k^\prime}}W_{r}^{\mu\nu}(q,p,S,k^\prime).
\end{align}
It is convenient to consider the  $k_\perp^\prime$-dependent cross section, i.e.,
\begin{align}
  d\sigma = \frac{\alpha_{\rm{em}}^2}{sQ^4}A_{r} L^r_{\mu\nu}(l,\lambda_e, l^\prime)W_r^{\mu\nu}(q,p,S,k_\perp^\prime) \frac{d^3 l^\prime d^2 k_\perp^\prime}{E_{l^\prime}}, \label{f:crosssection}
\end{align}
where the $k^{\prime}_z$ integrated TMD semi-inclusive hadronic tensor is given by
\begin{align}
W_r^{\mu\nu}(q,p,S,k_\perp^\prime) = \int \frac{dk_z^\prime}{(2\pi)^3 2E_{k^\prime}} W_{r}^{\mu\nu}(q,p,S,k^\prime).
\end{align}

In terms of the variables in Eq. (\ref{eq:SIDIS-var}), we have
\begin{align}
\frac{d^3 l^\prime}{2E_{l^\prime}} = \frac{y (s-M^2)}{4} dx d y d\psi \approx \frac{y s}{4} dx d y d\psi,
\end{align}
where $\psi$ is the azimuthal angle of $\vec l^\prime$ around $\vec l$, $M$ is the target mass which will be often neglected at high energy limit.
Therefore the cross section can be written as
\begin{align}
\frac{d\sigma}{dx dy d\psi d^2 k_\perp^\prime} = \frac{y \alpha_{\rm em}^2}{2 Q^4} \sum_r A_{r}
L_{\mu\nu}^r(l,\lambda_e, l^\prime)
W_r^{\mu\nu}(q,p,S,k_\perp^\prime).
\label{eq:Xsec-PartonModel}
\end{align}

\subsection{The general form of the cross section in terms of structure functions}

In considering the polarized reactions, the general form of the hadronic tensor is divided into a symmetric and an antisymmetric part,
$W^{\mu\nu} = W^{S\mu\nu}+ iW^{A\mu\nu}$,
where we have omitted the subscript $r=\gamma \gamma, ZZ, \gamma Z$ for simplicity. Furthermore, we have
\begin{align}
   W^{S\mu\nu} &=\sum_{\sigma,j} W_{\sigma j}^S h_{\sigma j}^{S\mu\nu} + \sum_{\sigma, j} \tilde W_{\sigma j}^S \tilde h_{\sigma j}^{S\mu\nu},\label{f:Wsmunu}\\
   W^{A\mu\nu} &=\sum_{\sigma,j} W_{\sigma j}^A h_{\sigma j}^{A\mu\nu} + \sum_{\sigma, j} \tilde W_{\sigma j}^A \tilde h_{\sigma j}^{A\mu\nu},\label{f:Wamunu}
\end{align}
where $h^{\mu\nu}_{\sigma j}$'s and $\tilde h^{\mu\nu}_{\sigma j}$'s represent the space reflection even and odd basic Lorentz tensors (BLTs), respectively.
They are constructed from available kinematical variables in the reaction process.
The subscript $\sigma$ specifies the polarizations.

It has been shown that a distinct feature for BLTs in semi-inclusive reactions is that
the polarization dependent BLTs can be taken as a product of the unpolarized BLTs and polarization dependent Lorentz scalar(s) or pseudo-scalar(s),
see \cite{Chen:2016moq} for the detailed discussions about the description of polarizations for spin-1 hadron and the construction of BLTs.
We repeat the results here for completeness.
There are 9 unpolarized BLTs given by
\begin{align}
   h^{S\mu\nu}_{Ui}&=\Big\{g^{\mu\nu}-\frac{q^\mu q^\nu}{q^2}, ~p_q^\mu  p_q^\nu,~ k_{q}^{\prime\mu}  k_{q}^{\prime\nu},  ~ p_q^{\{\mu} k_{ q}^{\prime\nu\}}\Big\},\label{f:hsU} \\
   \tilde h^{S\mu\nu}_{Ui}&=\Big\{\varepsilon^{\{\mu q p k^\prime } p_q^{\nu\}},~ \varepsilon^{\{\mu q p k^\prime }k_{ q}^{\prime\nu\}}\Big\}, \label{f:thsU} \\
   h^{A\mu\nu}_{U}&=\Big\{p_q^{[\mu} k_{q}^{\prime\nu]}\Big\},\label{f:haU} \\
   \tilde h^{A\mu\nu}_{Ui}&=\Big\{\varepsilon ^{\mu\nu qp},~ \varepsilon ^{\mu\nu qk^\prime }\Big\}.\label{f:thaU}
\end{align}
The subscript $U$ denotes the unpolarized part, and $p_q\equiv p - q(p\cdot q)/q^2$ satisfies $p_q \cdot q$ = 0.
We have also used notations $A^{\{\mu}B^{\nu\}} \equiv A^\mu B^\nu +A^\nu B^\mu$, and $A^{[\mu}B^{\nu]} \equiv A^\mu B^\nu -A^\nu B^\mu$.

As mentioned above, the vector polarization dependent BLTs can be constructed from the unpolarized BLTs and be written as a unified form given by
\begin{align}
  h^{S\mu\nu}_{Vi} &=\Big\{[\lambda_h, (k^\prime _\perp\cdot S_T)]\tilde h^{S\mu\nu}_{Ui}, ~\varepsilon_\perp^{k^\prime S} h^{S\mu\nu}_{Uj}\Big\}, \label{f:hsV}\\
 \tilde h^{S\mu\nu}_{Vi} &=\Big\{[\lambda_h, (k^\prime _\perp\cdot S_T)]h^{S\mu\nu}_{Ui}, ~ \varepsilon_\perp^{k^\prime S} \tilde h^{S\mu\nu}_{Uj} \Big\}, \label{f:thsV}\\
  h^{A\mu\nu}_{Vi} &=\Big\{[\lambda_h, (k^\prime _\perp\cdot S_T)]\tilde h^{A\mu\nu}_{Ui},~ \varepsilon_\perp^{k^\prime S}  h^{A\mu\nu}_{U}\Big\},\label{f:haV}\\
  \tilde h^{A\mu\nu}_{Vi} &=\Big\{[\lambda_h, (k^\prime _\perp\cdot S_T)]h^{A\mu\nu}_{U},~ \varepsilon_\perp^{k^\prime S}  \tilde h^{A\mu\nu}_{Uj}\Big\},\label{f:thaV}
\end{align}
where~$\varepsilon_\perp^{k^\prime S}=\varepsilon_\perp^{\alpha\beta} k^\prime _{\perp\alpha} S_{T\beta}$, $\varepsilon_\perp^{\alpha\beta}=\varepsilon^{\mu\nu\alpha\beta}\bar n_\mu n_\nu$; $\lambda_h$ is the hadron helicity while $S_T$ is the transverse polarization component. There are 27 such vector polarized BLTs in total.

The tensor polarized part is composed of $S_{LL}$-, $S_{LT}$- and $S_{TT}$-dependent parts. There are 9 $S_{LL}$-dependent BLTs, they are given by
\begin{align}
&h_{LLi}^{S\mu\nu}=S_{LL} h^{S\mu\nu}_{Ui}, \\
&\tilde h_{LLi}^{S\mu\nu}=S_{LL} \tilde h^{S\mu\nu}_{Ui}, \label{f:hsLL} \\
&h_{LL}^{A\mu\nu}=S_{LL} h^{A\mu\nu}_{U}, \\
&\tilde h_{LLi}^{A\mu\nu}=S_{LL} \tilde h^{A\mu\nu}_{Ui}. \label{f:thsLL}
\end{align}
The $S_{LT}$ part can be obtained from Eqs.~(\ref{f:hsV})-(\ref{f:thaV}) with replacing $S_T$ by $S_{LT}$, i.e.,
\begin{align}
 & h^{S\mu\nu}_{LTi} =\Big\{ (k^\prime _\perp\cdot S_{LT}) h^{S\mu\nu}_{Ui}, ~\varepsilon_\perp^{k^\prime S_{LT}} \tilde h^{S\mu\nu}_{Uj}\Big\}, \label{f:hsLT}\\
 & \tilde h^{S\mu\nu}_{LTi} =\Big\{(k^\prime _\perp\cdot S_{LT})\tilde h^{S\mu\nu}_{Ui}, ~ \varepsilon_\perp^{k^\prime S_{LT}} h^{S\mu\nu}_{Uj} \Big\}, \label{f:thsLT}\\
 & h^{A\mu\nu}_{LTi} =\Big\{(k^\prime _\perp\cdot S_{LT}) h^{A\mu\nu}_{U},~ \varepsilon_\perp^{k^\prime S_{LT}}  \tilde h^{A\mu\nu}_{Uj}\Big\},\label{f:haLT}\\
 & \tilde h^{A\mu\nu}_{LTi} =\Big\{(k^\prime _\perp\cdot S_{LT})\tilde h^{A\mu\nu}_{Ui},~ \varepsilon_\perp^{k^\prime S_{LT}} h^{A\mu\nu}_{U}\Big\}. \label{f:thaLT}
\end{align}
For the $S_{TT}$ part, we have
\begin{align}
 & h^{S\mu\nu}_{TTi} =\Big\{ S_{TT}^{k^\prime k^\prime} h^{S\mu\nu}_{Ui}, ~\tilde S_{TT}^{k^\prime k^\prime}  \tilde h^{S\mu\nu}_{Uj}\Big\}, \label{f:hsTT}\\
 & \tilde h^{S\mu\nu}_{TTi} =\Big\{S_{TT}^{k^\prime k^\prime} \tilde h^{S\mu\nu}_{Ui}, ~ \tilde S_{TT}^{ k^\prime k^\prime} h^{S\mu\nu}_{Uj} \Big\}, \label{f:thsTT}\\
 & h^{A\mu\nu}_{TTi} =\Big\{S_{TT}^{k^\prime k^\prime}  h^{A\mu\nu}_{U},~ \tilde S_{TT}^{k^\prime k^\prime }  \tilde h^{A\mu\nu}_{Uj}\Big\}, \label{f:haTT}\\
 & \tilde h^{A,\mu\nu}_{TTi} =\Big\{S_{TT}^{k^\prime k^\prime} \tilde h^{A\mu\nu}_{Ui},~ \tilde S_{TT}^{ k^\prime k^\prime } h^{A\mu\nu}_{U}\Big\},\label{f:thaTT}
\end{align}
where~ $S_{TT}^{k^\prime k^\prime}=k^\prime _{\perp\alpha} S_{TT}^{\alpha\beta} k^\prime_{\perp\beta}$, $\tilde k_{\perp}^{\prime\alpha}=\varepsilon_{\perp}^{\alpha\beta} k^\prime_{\perp\beta}$.
There are 81 such BLTs in total. 

In expressing the cross section, we choose a coordinate system so that the momenta related to this SIDIS process take the following forms:
\begin{align}
& p^\mu = \left(p^+,0,\vec 0_\perp \right), \nonumber\\
& l^\mu = \left( \frac{1-y}{y}xp^+, \frac{Q^2}{2xyp^+}, \frac{Q\sqrt{1-y}}{y},0 \right),\nonumber\\
& q^\mu = \left( -xp^+, \frac{Q^2}{2xp^+}, \vec 0_\perp \right), \nonumber\\
& k_\perp^{\prime\mu} = k_\perp^\mu = |\vec k_\perp| \left( 0,0, \cos\varphi, \sin\varphi \right).
\end{align}
And the transverse vector polarization is parameterized as
\begin{align}
& S_T^\mu = |\vec S_T| \left( 0,0, \cos\varphi_S, \sin\varphi_S \right).
\end{align}
For the tensor polarization dependent parameters, we parameterize and define them as in Ref.~\cite{Bacchetta:2000jk}, i.e.,
\begin{align}
& S_{LT}^x = |S_{LT}| \cos\varphi_{LT}, \\
& S_{LT}^y = |S_{LT}| \sin\varphi_{LT}, \\
& |S_{LT}| = \sqrt{(S_{LT}^x)^2 + (S_{LT}^y)^2}, \\
& S_{TT}^{xx} = -S_{TT}^{yy} = |S_{TT}| \cos2\varphi_{TT}, \\
& S_{TT}^{xy} = S_{TT}^{yx} = |S_{TT}| \sin2\varphi_{TT}, \\
& |S_{TT}| = \sqrt{(S_{TT}^{xx})^2 + (S_{TT}^{xy})^2}.
\end{align}

After making Lorentz contractions with the leptonic tensor, we obtain the general form for the cross section. We give the general form of the cross section through weak interaction channel. The cross section is given by
\begin{align}
  \frac{d\sigma^{ZZ}}{dxdyd\psi d^2k^\prime _\perp}&=\frac{\alpha_{\rm em}^2}{yQ^2}\chi \Big[ \mathcal{W}_{U,U}+\lambda_e\mathcal{W}_{L,U}\nonumber\\
  &+ \lambda_h\mathcal{W}_{U,L} + \lambda_e\lambda_h\mathcal{W}_{L,L}\nonumber\\
  &+S_{LL}\mathcal{W}_{U,LL} + \lambda_eS_{LL}\mathcal{W}_{L,LL}\nonumber\\
  &+|S_T|\mathcal{W}_{U,T} + \lambda_e|S_T|\mathcal{W}_{L,T}\nonumber\\
  &+|S_{LT}|\mathcal{W}_{U,LT} + \lambda_e|S_{LT}|\mathcal{W}_{L,LT}\nonumber\\
  &+|S_{TT}|\mathcal{W}_{U,TT} + \lambda_e|S_{TT}|\mathcal{W}_{L,TT}\Big],
\end{align}
The total cross section including the electromagnetic and interference terms will formally take the same structure. We define functions of $y$ which will be often used:
\begin{align}
& A(y) = y^2-2y+2, \nonumber\\
& B(y) = 2(2-y)\sqrt{1-y}, \nonumber\\
& C(y) = y(2-y), \nonumber\\
& D(y) = 2y\sqrt{1-y}, \nonumber\\
& E(y) = 2(1-y).
\end{align}
In expressing the cross section, these functions are equivalent to variables $\cal K$ and $\varepsilon$ used in Ref.~\cite{Bacchetta:2006tn}.
From $\varepsilon=(1-y-\frac{1}{4}\gamma^2y^2)/(1-y+\frac{1}{2}y^2+\frac{1}{4}\gamma^2y^2)$ and ${\cal K}=(1+{\gamma^2}/{2x}) {y^2}/{(1-\varepsilon)}$ with $\gamma=2Mx/Q$, if neglecting the hadron mass, i.e., $\gamma = 0$, we have
\begin{align}
& {\cal K} = A(y), \nonumber\\
& {\cal K} \varepsilon = E(y), \nonumber\\
& {\cal K} \sqrt{2\varepsilon(1+\varepsilon)} = B(y), \nonumber\\
& {\cal K} \sqrt{2\varepsilon(1-\varepsilon)} = D(y), \nonumber\\
& {\cal K} \sqrt{1-\varepsilon^2} = C(y).
\end{align}
The explicit results for the cross section in terms of structure functions for different polarization configurations are:
\begin{widetext}
\begin{align}
{\cal W}_{U,U} &= A(y) W_{U,U}^T + E(y) W_{U,U}^L + B(y)\left( \sin\varphi \tilde W_{U,U1}^{\sin\varphi} + \cos\varphi W_{U,U1}^{\cos\varphi} \right) + E(y)\left( \sin2\varphi \tilde W_{U,U}^{\sin2\varphi} + \cos2\varphi W_{U,U}^{\cos2\varphi} \right)  \nonumber\\
& + C(y) W_{U,U} + D(y) \left( \sin\varphi \tilde W_{U,U2}^{\sin\varphi} + \cos\varphi W_{U,U2}^{\cos\varphi} \right), \\
{\cal W}_{L,U} &= A(y) \tilde W_{L,U}^T + E(y) \tilde W_{L,U}^L + B(y)\left( \sin\varphi W_{L,U1}^{\sin\varphi} + \cos\varphi \tilde W_{L,U1}^{\cos\varphi} \right) + E(y)\left( \sin2\varphi W_{L,U}^{\sin2\varphi} + \cos2\varphi \tilde W_{L,U}^{\cos2\varphi} \right) \nonumber\\
& + C(y) \tilde W_{L,U} + D(y) \left( \sin\varphi W_{L,U2}^{\sin\varphi} + \cos\varphi \tilde W_{L,U2}^{\cos\varphi} \right), \\
{\cal W}_{U,L} &= A(y) \tilde W_{U,L}^T + E(y) \tilde W_{U,L}^L + B(y)\left( \sin\varphi W_{U,L1}^{\sin\varphi} + \cos\varphi \tilde W_{U,L1}^{\cos\varphi} \right) + E(y)\left( \sin2\varphi W_{U,L}^{\sin2\varphi} + \cos2\varphi \tilde W_{U,L}^{\cos2\varphi} \right)  \nonumber\\
& + C(y) \tilde W_{U,L} + D(y) \left( \sin\varphi W_{U,L2}^{\sin\varphi} + \cos\varphi \tilde W_{U,L2}^{\cos\varphi} \right), \\
{\cal W}_{L,L} &= A(y) W_{L,L}^T + E(y) W_{L,L}^L + B(y)\left( \sin\varphi \tilde W_{L,L1}^{\sin\varphi} + \cos\varphi W_{L,L1}^{\cos\varphi} \right) + E(y)\left( \sin2\varphi \tilde W_{L,L}^{\sin2\varphi} + \cos2\varphi W_{L,L}^{\cos2\varphi} \right)  \nonumber\\
& + C(y) W_{L,L} + D(y) \left( \sin\varphi \tilde W_{L,L2}^{\sin\varphi} + \cos\varphi W_{L,L2}^{\cos\varphi} \right), \\
{\cal W}_{U,LL} &= A(y) W_{U,LL}^T + E(y) W_{U,LL}^L + B(y)\left( \sin\varphi \tilde W_{U,LL1}^{\sin\varphi} + \cos\varphi W_{U,LL1}^{\cos\varphi} \right) + E(y)\left( \sin2\varphi \tilde W_{U,LL}^{\sin2\varphi} + \cos2\varphi W_{U,LL}^{\cos2\varphi} \right)  \nonumber\\
& + C(y) W_{U,LL} + D(y) \left( \sin\varphi \tilde W_{U,LL2}^{\sin\varphi} + \cos\varphi W_{U,LL2}^{\cos\varphi} \right), \\
{\cal W}_{L,LL} &= A(y) \tilde W_{L,LL}^T + E(y) \tilde W_{L,LL}^L + B(y)\left( \sin\varphi W_{L,LL1}^{\sin\varphi} + \cos\varphi \tilde W_{L,LL1}^{\cos\varphi} \right) + E(y)\left( \sin2\varphi W_{L,LL}^{\sin2\varphi} + \cos2\varphi \tilde W_{L,LL}^{\cos2\varphi} \right) \nonumber\\
& + C(y) \tilde W_{L,LL} + D(y) \left( \sin\varphi W_{L,LL2}^{\sin\varphi} + \cos\varphi \tilde W_{L,LL2}^{\cos\varphi} \right), \\
{\cal W}_{U,T} &= \sin\varphi_S \left[ B(y) W_{U,T1}^{\sin\varphi_S} + D(y) W_{U,T2}^{\sin\varphi_S} \right]
+ \sin(\varphi+\varphi_S) E(y) W_{U,T}^{\sin(\varphi+\varphi_S)} \nonumber\\
&+ \sin(\varphi-\varphi_S) \left[ A(y) W_{U,T}^{T,\sin(\varphi-\varphi_S)} + E(y) W_{U,T}^{L,\sin(\varphi-\varphi_S)} + C(y) W_{U,T}^{\sin(\varphi-\varphi_S)} \right] \nonumber\\
&+ \sin(2\varphi-\varphi_S) \left[ B(y) W_{U,T1}^{\sin(2\varphi-\varphi_S)} + D(y) W_{U,T2}^{\sin(2\varphi-\varphi_S)} \right]
+ \sin(3\varphi-\varphi_S) E(y) W_{U,T}^{\sin(3\varphi-\varphi_S)} \nonumber\\
&+ \cos\varphi_S \left[ B(y) \tilde W_{U,T1}^{\cos\varphi_S} + D(y) \tilde W_{U,T2}^{\cos\varphi_S} \right]
+ \cos(\varphi+\varphi_S) E(y) \tilde W_{U,T}^{\cos(\varphi+\varphi_S)} \nonumber\\
&+ \cos(\varphi-\varphi_S) \left[ A(y) \tilde W_{U,T}^{T,\cos(\varphi-\varphi_S)} + E(y) \tilde W_{U,T}^{L,\cos(\varphi-\varphi_S)} + C(y) \tilde W_{U,T}^{\cos(\varphi-\varphi_S)} \right] \nonumber\\
&+ \cos(2\varphi-\varphi_S) \left[ B(y) \tilde W_{U,T1}^{\cos(2\varphi-\varphi_S)} + D(y) \tilde W_{U,T2}^{\cos(2\varphi-\varphi_S)} \right]
+ \cos(3\varphi-\varphi_S) E(y) \tilde W_{U,T}^{\cos(3\varphi-\varphi_S)}, \\
{\cal W}_{L,T} &= \sin\varphi_S \left[ B(y) \tilde W_{L,T1}^{\sin\varphi_S} + D(y) \tilde W_{L,T2}^{\sin\varphi_S} \right]
+ \sin(\varphi+\varphi_S) E(y) \tilde W_{L,T}^{\sin(\varphi+\varphi_S)} \nonumber\\
&+ \sin(\varphi-\varphi_S) \left[ A(y) \tilde W_{L,T}^{T,\sin(\varphi-\varphi_S)} + E(y) \tilde W_{L,T}^{L,\sin(\varphi-\varphi_S)} + C(y) \tilde W_{L,T}^{\sin(\varphi-\varphi_S)} \right] \nonumber\\
&+ \sin(2\varphi-\varphi_S) \left[ B(y) \tilde W_{L,T1}^{\sin(2\varphi-\varphi_S)} + D(y) \tilde W_{L,T2}^{\sin(2\varphi-\varphi_S)} \right]
+ \sin(3\varphi-\varphi_S) E(y) \tilde W_{L,T}^{\sin(3\varphi-\varphi_S)} \nonumber\\
&+ \cos\varphi_S \left[ B(y) W_{L,T1}^{\cos\varphi_S} + D(y) W_{L,T2}^{\cos\varphi_S} \right]
+ \cos(\varphi+\varphi_S) E(y) W_{L,T}^{\cos(\varphi+\varphi_S)} \nonumber\\
&+ \cos(\varphi-\varphi_S) \left[ A(y) W_{L,T}^{T,\cos(\varphi-\varphi_S)} + E(y) W_{L,T}^{L,\cos(\varphi-\varphi_S)} + C(y) W_{L,T}^{\cos(\varphi-\varphi_S)} \right] \nonumber\\
&+ \cos(2\varphi-\varphi_S) \left[ B(y) W_{L,T1}^{\cos(2\varphi-\varphi_S)} + D(y) W_{L,T2}^{\cos(2\varphi-\varphi_S)} \right]
+ \cos(3\varphi-\varphi_S) E(y) W_{L,T}^{\cos(3\varphi-\varphi_S)},  \\
{\cal W}_{U,LT} &= \sin\varphi_{LT} \left[ B(y) \tilde W_{U,LT1}^{\sin\varphi_{LT}} + D(y) \tilde W_{U,LT2}^{\sin\varphi_{LT}} \right]
+ \sin(\varphi+\varphi_{LT}) E(y) \tilde W_{U,LT}^{\sin(\varphi+\varphi_{LT})} \nonumber\\
&+ \sin(\varphi-\varphi_{LT}) \left[ A(y) \tilde W_{U,LT}^{T,\sin(\varphi-\varphi_{LT})} + E(y) \tilde W_{U,LT}^{L,\sin(\varphi-\varphi_{LT})} + C(y) \tilde W_{U,LT}^{\sin(\varphi-\varphi_{LT})} \right] \nonumber\\
&+ \sin(2\varphi-\varphi_{LT}) \left[ B(y) \tilde W_{U,LT1}^{\sin(2\varphi-\varphi_{LT})} + D(y) \tilde W_{U,LT2}^{\sin(2\varphi-\varphi_{LT})} \right]
+ \sin(3\varphi-\varphi_{LT}) E(y) \tilde W_{U,LT}^{\sin(3\varphi-\varphi_{LT})} \nonumber\\
&+ \cos\varphi_{LT} \left[ B(y) W_{U,LT1}^{\cos\varphi_{LT}} + D(y) W_{U,LT2}^{\cos\varphi_{LT}} \right]
+ \cos(\varphi+\varphi_{LT}) E(y) W_{U,LT}^{\cos(\varphi+\varphi_{LT})} \nonumber\\
&+ \cos(\varphi-\varphi_{LT}) \left[ A(y) W_{U,LT}^{T,\cos(\varphi-\varphi_{LT})} + E(y) W_{U,LT}^{L,\cos(\varphi-\varphi_{LT})} + C(y) W_{U,LT}^{\cos(\varphi-\varphi_{LT})} \right] \nonumber\\
&+ \cos(2\varphi-\varphi_{LT}) \left[ B(y) W_{U,LT1}^{\cos(2\varphi-\varphi_{LT})} + D(y) W_{U,LT2}^{\cos(2\varphi-\varphi_{LT})} \right]
+ \cos(3\varphi-\varphi_{LT}) E(y) W_{U,LT}^{\cos(3\varphi-\varphi_{LT})}, \\
{\cal W}_{L,LT} &= \sin\varphi_{LT} \left[ B(y) W_{L,LT1}^{\sin\varphi_{LT}} + D(y) W_{L,LT2}^{\sin\varphi_{LT}} \right]
+ \sin(\varphi+\varphi_{LT}) E(y) W_{L,LT}^{\sin(\varphi+\varphi_{LT})} \nonumber\\
&+ \sin(\varphi-\varphi_{LT}) \left[ A(y) W_{L,LT}^{T,\sin(\varphi-\varphi_{LT})} + E(y) W_{L,LT}^{L,\sin(\varphi-\varphi_{LT})} + C(y) W_{L,LT}^{\sin(\varphi-\varphi_{LT})} \right] \nonumber\\
&+ \sin(2\varphi-\varphi_{LT}) \left[ B(y) W_{L,LT1}^{\sin(2\varphi-\varphi_{LT})} + D(y) W_{L,LT2}^{\sin(2\varphi-\varphi_{LT})} \right]
+ \sin(3\varphi-\varphi_{LT}) E(y) W_{L,LT}^{\sin(3\varphi-\varphi_{LT})} \nonumber\\
&+ \cos\varphi_{LT} \left[ B(y) \tilde W_{L,LT1}^{\cos\varphi_{LT}} + D(y) \tilde W_{L,LT2}^{\cos\varphi_{LT}} \right]
+ \cos(\varphi+\varphi_{LT}) E(y) \tilde W_{L,LT}^{\cos(\varphi+\varphi_{LT})} \nonumber\\
&+ \cos(\varphi-\varphi_{LT}) \left[ A(y) \tilde W_{L,LT}^{T,\cos(\varphi-\varphi_{LT})} + E(y) \tilde W_{L,LT}^{L,\cos(\varphi-\varphi_{LT})} + C(y) \tilde W_{L,LT}^{\cos(\varphi-\varphi_{LT})} \right] \nonumber\\
&+ \cos(2\varphi-\varphi_{LT}) \left[ B(y) \tilde W_{L,LT1}^{\cos(2\varphi-\varphi_{LT})} + D(y) \tilde W_{L,LT2}^{\cos(2\varphi-\varphi_{LT})} \right]
+ \cos(3\varphi-\varphi_{LT}) E(y) \tilde W_{L,LT}^{\cos(3\varphi-\varphi_{LT})}, \\
{\cal W}_{U,TT} &= \sin(\varphi-2\varphi_{TT}) \left[ B(y) \tilde W_{U,TT1}^{\sin(\varphi-2\varphi_{TT})} + D(y) \tilde W_{U,TT2}^{\sin(\varphi-2\varphi_{TT})} \right]
+ \sin2\varphi_{TT} E(y) \tilde W_{U,TT}^{\sin2\varphi_{TT}} \nonumber\\
&+ \sin(2\varphi-2\varphi_{TT}) \left[ A(y) \tilde W_{U,TT}^{T,\sin(2\varphi-2\varphi_{TT})} + E(y) \tilde W_{U,TT}^{L,\sin(2\varphi-2\varphi_{TT})} + C(y) \tilde W_{U,TT}^{\sin(2\varphi-2\varphi_{TT})} \right] \nonumber\\
&+ \sin(3\varphi-2\varphi_{TT}) \left[ B(y) \tilde W_{U,TT1}^{\sin(3\varphi-2\varphi_{TT})} + D(y) \tilde W_{U,TT2}^{\sin(3\varphi-2\varphi_{TT})} \right]
+ \sin(4\varphi-2\varphi_{TT}) E(y) \tilde W_{U,TT}^{\sin(4\varphi-2\varphi_{TT})} \nonumber\\
&+ \cos(\varphi-2\varphi_{TT}) \left[ B(y) W_{U,TT1}^{\cos(\varphi-2\varphi_{TT})} + D(y) W_{U,TT2}^{\cos(\varphi-2\varphi_{TT})} \right]
+ \cos2\varphi_{TT} E(y) W_{U,TT}^{\cos2\varphi_{TT}} \nonumber\\
&+ \cos(2\varphi-2\varphi_{TT}) \left[ A(y) W_{U,TT}^{T,\cos(2\varphi-2\varphi_{TT})} + E(y) W_{U,TT}^{L,\cos(2\varphi-2\varphi_{TT})} + C(y) W_{U,TT}^{\cos(2\varphi-2\varphi_{TT})} \right] \nonumber\\
&+ \cos(3\varphi-2\varphi_{TT}) \left[ B(y) W_{U,TT1}^{\cos(3\varphi-2\varphi_{TT})} + D(y) W_{U,TT2}^{\cos(3\varphi-2\varphi_{TT})} \right]
+ \cos(4\varphi-2\varphi_{TT}) E(y) W_{U,TT}^{\cos(4\varphi-2\varphi_{TT})}, \\
{\cal W}_{L,TT} &= \sin(\varphi-2\varphi_{TT}) \left[ B(y) W_{L,TT1}^{\sin(\varphi-2\varphi_{TT})} + D(y) W_{L,TT2}^{\sin(\varphi-2\varphi_{TT})} \right]
+ \sin2\varphi_{TT} E(y) W_{L,TT}^{\sin2\varphi_{TT}} \nonumber\\
&+ \sin(2\varphi-2\varphi_{TT}) \left[ A(y) W_{L,TT}^{T,\sin(2\varphi-2\varphi_{TT})} + E(y) W_{L,TT}^{L,\sin(2\varphi-2\varphi_{TT})} + C(y) W_{L,TT}^{\sin(2\varphi-2\varphi_{TT})} \right] \nonumber\\
&+ \sin(3\varphi-2\varphi_{TT}) \left[ B(y) W_{L,TT1}^{\sin(3\varphi-2\varphi_{TT})} + D(y) W_{L,TT2}^{\sin(3\varphi-2\varphi_{TT})} \right]
+ \sin(4\varphi-2\varphi_{TT}) E(y) W_{L,TT}^{\sin(4\varphi-2\varphi_{TT})} \nonumber\\
&+ \cos(\varphi-2\varphi_{TT}) \left[ B(y) \tilde W_{L,TT1}^{\cos(\varphi-2\varphi_{TT})} + D(y) \tilde W_{L,TT2}^{\cos(\varphi-2\varphi_{TT})} \right]
+ \cos2\varphi_{TT} E(y) \tilde W_{L,TT}^{\cos2\varphi_{TT}} \nonumber\\
&+ \cos(2\varphi-2\varphi_{TT}) \left[ A(y) \tilde W_{L,TT}^{T,\cos(2\varphi-2\varphi_{TT})} + E(y) \tilde W_{L,TT}^{L,\cos(2\varphi-2\varphi_{TT})} + C(y) \tilde W_{L,TT}^{\cos(2\varphi-2\varphi_{TT})} \right] \nonumber\\
&+ \cos(3\varphi-2\varphi_{TT}) \left[ B(y) \tilde W_{L,TT1}^{\cos(3\varphi-2\varphi_{TT})} + D(y) \tilde W_{L,TT2}^{\cos(3\varphi-2\varphi_{TT})} \right]
+ \cos(4\varphi-2\varphi_{TT}) E(y) \tilde W_{L,TT}^{\cos(4\varphi-2\varphi_{TT})}.
\end{align}
For both unpolarized and polarized electron cases, there exist 81 structure functions respectively, which correspond to the number of independent BLTs. For unpolarized electron beam, there are 39 structure functions correspond to parity conserved terms and the other 42 are parity violated. While for polarized electron beam, it's just opposite, i.e., 42 structure functions are parity conserved and 39 are parity violated.
\end{widetext}

\section{The hadronic tensor in the QCD parton model} \label{sec:partonmodel}
\subsection{The collinear expansion}
\begin{figure} [htb]
\centering
\includegraphics[width= 0.95\linewidth]{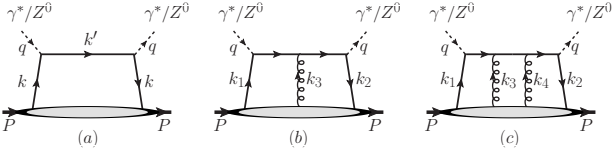}
\caption{The first few diagrams of the Feynman diagram series with exchange of $j$ gluons, where $j=0,~1$ and $2$ for diagrams $(a)$, $(b)$ and $(c)$ respectively.}
\label{fig:CoExp}
\end{figure}

In the QCD parton model, we can calculate the hadronic tensor in terms of gauge-invariant TMD PDFs.
At the tree level, we need to consider the contributions from the series of diagrams shown in Fig.~\ref{fig:CoExp}, i.e., the multiple gluon scattering contributions.

After collinear expansion, the hadronic tensor is expressed in terms of
the gauge-invariant quark-quark and quark-j-gluon(s)-quark correlators and calculable hard parts \cite{Liang:2006wp,Song:2010pf,Song:2013sja},
\begin{align}
W_{r,\mu\nu} (q,p,S,k^\prime) = \sum_{j,c} \tilde W_{r,\mu\nu}^{(j,c)} (q,p,S,k^\prime),
\end{align}
where $j$ denotes the number of gluons exchanged and $c$ denotes different cuts. After integration over $k_z^\prime$, $\tilde W_{r,\mu\nu}^{(j,c)}$'s are simplified as
\begin{align}
& \tilde{W}_{r,\mu\nu}^{(0)}(q,p,S,k_\perp^\prime) = \frac{1}{2}{\rm Tr}\left[\hat{h}_{r,\mu \nu}^{(0)} \hat{\Phi}^{(0)}(x,k_\perp^\prime)\right] , \label{f:W0munu}\\
& \tilde{W}_{r,\mu\nu}^{(1, L)}(q,p,S,k_\perp^\prime) = \frac{1}{4 p \cdot q}{\rm Tr}\left[\hat{h}_{r,\mu\nu}^{(1) \rho} \hat{\varphi}_{\rho}^{(1)}(x,k_\perp^\prime)\right], \label{f:W1Lmunu}
\end{align}
up to the relevant twist-3 level. The hard parts $h_r$'s are
\begin{align}
& \hat h_{\gamma\gamma,\mu\nu}^{(0)} = \gamma_{\mu} \slashed n \gamma_{\nu} / p^+, \qquad \hat{h}_{\gamma\gamma,\mu\nu}^{(1) \rho}=\gamma_{\mu} \slashed{\bar n} \gamma_{\perp}^{\rho} \slashed n \gamma_{\nu}, \\
& \hat h_{ZZ,\mu\nu}^{(0)} = \Gamma_{\mu}^q \slashed n \Gamma_{\nu}^q / p^+, \qquad \hat{h}_{ZZ,\mu\nu}^{(1) \rho}=\Gamma_{\mu}^q \slashed{\bar n} \gamma_{\perp}^{\rho} \slashed n \Gamma_{\nu}^q, \\
& \hat h_{\gamma Z,\mu\nu}^{(0)} = \Gamma_{\mu}^q \slashed n \gamma_{\nu} / p^+, \qquad \hat{h}_{\gamma Z,\mu\nu}^{(1) \rho}=\Gamma_{\mu}^q \slashed{\bar n} \gamma_{\perp}^{\rho} \slashed n \gamma_{\nu}.
\end{align}

The gauge-invariant quark-quark and quark-gluon-quark correlators are defined as
\begin{align}
  \hat{\Phi}^{(0)}\left(x, k_{\perp}\right) =& \int \frac{p^{+} d y^{-} d^{2} y_{\perp}}{(2 \pi)^{3}} e^{i x p^{+} y^{-}-i \vec{k}_{\perp} \cdot \vec{y}_{\perp}} \nonumber\\
  &\times \langle N|\bar{\psi}(0) {\cal L}(0, y) \psi(y)| N\rangle, \\
  \hat{\varphi}_{\rho}^{(1)}\left(x, k_{\perp}\right) =& \int \frac{p^{+} d y^{-} d^{2} y_{\perp}}{(2 \pi)^{3}} e^{i x p^+ y^- - i \vec{k}_\perp \cdot \vec{y}_\perp}\nonumber\\
  &\times  \langle N | \bar{\psi}(0) D_{\perp \rho}(0) {\cal L}(0, y) \psi(y)| N\rangle,
\end{align}
where $D_\rho(y) = -i\partial_\rho + g A_\rho(y)$ is the covariant derivative.
${\cal L}(0, y)$ is the gauge link obtained from the collinear expansion procedure, which guarantees the gauge invariance of the correlators.

\subsection{Decomposition of quark-quark and quark-gluon-quark correlators}

The quark-quark and quark-gluon-quark correlators are $4\times 4$ matrices in Dirac space which can be decomposed in terms of the Dirac Gamma-matrices $\{I, i\gamma^5, \gamma^\alpha, \gamma^\alpha\gamma^5, i\sigma^{\alpha\beta}\gamma^5\}$.
In the SIDIS process $e^-N\to e^- q X$,  where the fragmentation is not considered, only the chiral even PDFs are involved.
Thus we only need to consider the $\gamma^\alpha$- and the $\gamma^\alpha\gamma^5$-terms in the decomposition of the correlators.
We have
\begin{align}
& \hat \Phi^{(0)}= \frac{1}{2}\left[\gamma^\alpha \Phi^{(0)}_\alpha + \gamma^\alpha\gamma_5 \tilde\Phi^{(0)}_\alpha \right] + \cdots, \\
& \hat \varphi_\rho^{(1)}= \frac{1}{2}\left[\gamma^\alpha \varphi_{\rho\alpha}^{(1)} + \gamma^\alpha\gamma_5 \tilde\varphi_{\rho\alpha}^{(1)} \right] + \cdots.
\end{align}

The TMD PDFs are defined through the decomposition of the correlation functions.
Following the convention in Ref.~\cite{Wei:2016far}, we have
\begin{align}
  \Phi^{(0)}_\alpha &=p^+ \bar n_\alpha\Bigl(f_1+S_{LL}f_{1LL}-\frac{k_\perp \cdot \tilde S_T}{M}f^\perp_{1T} \nonumber\\ &+\frac{k_\perp \cdot S_{LT}}{M}f_{1LT}^{\perp}+\frac{S_{TT}^{kk}}{M^2}f_{1TT}^{\perp} \Bigr) +k_{\perp\alpha}\Big( f^\perp +S_{LL}f_{LL}^\perp  \Big)  \nonumber\\
  &- M\tilde S_{T\alpha}f_T + M S_{LT\alpha}f_{LT}+ S^k_{TT\alpha} f_{TT} - \lambda_h \tilde k_{\perp\alpha} f_L^\perp \nonumber\\ & -\frac{k_{\perp\langle\alpha}k_{\perp\beta\rangle}}{M} \Bigl( \tilde S_T^\beta f_T^\perp + S_{LT}^\beta f_{LT}^\perp + \frac{ S_{TT}^{k\beta}}{M} f_{TT}^\perp \Bigr),
\label{eq:Xi0Peven}\\
  \tilde\Phi^{(0)}_\alpha &=p^+\bar n_\alpha\Bigl(-\lambda_hg_{1L}+\frac{k_\perp\cdot S_T}{M}g^\perp_{1T}\nonumber\\
  &+ \frac{k_\perp \cdot \tilde S_{LT}}{M} g_{1LT}^\perp - \frac{\tilde S_{TT}^{k k}}{M^2} g_{1TT}^\perp\Bigr)-\tilde k_{\perp\alpha} \Big( g^\perp+S_{LL}g_{LL}^\perp \Big) \nonumber\\
  &- M S_{T\alpha}g_T - M \tilde S_{LT\alpha}g_{LT} - \tilde S^{k}_{TT\alpha} g_{TT} -\lambda_h k_{\perp\alpha} g_L^\perp \nonumber\\
  &+ \frac{k_{\perp\langle\alpha}k_{\perp\beta\rangle}}{M} \Bigl( S_T^\beta g_T^\perp - \tilde S_{LT}^\beta g_{LT}^\perp - \frac{ S_{TT}^{k\beta}}{M} \tilde g_{TT}^\perp \Bigr).
\label{eq:Xi0Podd}
\end{align}
 For the quark-gluon-quark correlator, we have
\begin{align}
  \varphi^{(1)}_{\rho\alpha}&=p^+\bar n_\alpha\Biggl[k_{\perp\rho}\big( f^\perp_d+ S_{LL}f_{dLL}^\perp\big)- M\tilde S_{T\rho}f_{dT} \nonumber\\
  & +MS_{LT\rho} f_{dLT}+S_{TT\rho}^k f_{dTT} -\lambda_h \tilde k_{\perp\rho} f_{dL}^\perp \nonumber\\
  & -\frac{k_{\perp\langle\rho}k_{\perp\beta\rangle}}{M} \Bigl( \tilde S_T^\beta f_{dT}^\perp + S_{LT}^\beta f_{dLT}^\perp + \frac{S_{TT}^{k\beta}}{M} f_{dTT}^\perp \Bigr)\Biggr], \label{eq:Xi1Peven} \\
  \tilde \varphi^{(1)}_{\rho\alpha}&=ip^+\bar n_\alpha\Biggl[\tilde k_{\perp\rho}\big( g^\perp_d+ S_{LL}g_{dLL}^\perp\big) + MS_{T\rho}g_{dT}\nonumber\\
  & + M\tilde S_{LT\rho} g_{dLT} + \tilde S_{TT\rho}^{k} g_{dTT}+\lambda_h k_{\perp\rho} g_{dL}^\perp  \nonumber\\
  & - \frac{k_{\perp\langle\rho}k_{\perp\beta\rangle}}{M} \Bigl( S_T^\beta g_{dT}^\perp - \tilde S_{LT}^\beta g_{dLT}^\perp - \frac{\tilde S_{TT}^{k\beta}}{M} g_{dTT}^\perp \Bigr) \Biggr],
\label{eq:Xi1Podd}
\end{align}
where
$S_{TT}^{k\beta} \equiv S_{TT}^{\alpha\beta}k_{\perp\alpha}$, and $\tilde S_{TT}^{k\beta} \equiv \varepsilon_{\perp\mu}^{\beta} S_{TT}^{k\mu}$. We have $\frac{1}{M} S_{TT}^{k\beta}$ behaves as a Lorentz vector like $S_{LT}^\beta$, $\frac{1}{M} \tilde S_{TT}^{k\beta}$ and $\tilde S_{LT}^\beta$ behave as axial vectors like $S_{T}^\beta$.

In fact, not all of these TMD PFDs shown in Eqs.~(\ref{eq:Xi0Peven})-(\ref{eq:Xi1Podd}) are independent. We use the QCD equation of motion $\slashed D\psi=0$ to obtain the following equations to eliminate PDFs which are not independent, i.e.,
\begin{align}
 x p^{+} \Phi^{(0) \rho} &=-g_{\perp}^{\rho \sigma} \operatorname{Re} \varphi_{\sigma+}^{(1)}-\varepsilon_{\perp}^{\rho \sigma} \operatorname{Im} \tilde{\varphi}_{\sigma+}^{(1)}, \label{eq:eom1}\\
 x p^{+} \tilde{\Phi}^{(0) \rho} &=-g_{\perp}^{\rho \sigma} \operatorname{Re} \tilde{\varphi}_{\sigma+}^{(1)}-\varepsilon_{\perp}^{\rho \sigma} \operatorname{Im} \varphi_{\sigma+}^{(1)}.\label{eq:eom2}
\end{align}
By inserting Eqs.~(\ref{eq:Xi0Peven})-(\ref{eq:Xi1Podd}) into Eqs.~(\ref{eq:eom1}) and (\ref{eq:eom2}), we can get the relationships between the twist-3 TMD PDFs defined via the quark-quark correlator and those defined via the quark-gluon-quark correlator.
They can be written in a unified form, i.e.,
\begin{align}
f_{d S}^{K}-g_{d S}^{K}=-x\left(f_{S}^{K}-i g_{S}^{K}\right),\label{f:formulaEOM}
\end{align}
where $k=$null, $\perp$, $S=$null, $L$, $T$, $LL$, $LT$ and $TT$ whenever applicable.

\subsection{The hadronic tensor results}

Substituting the Lorentz decomposition expressions of the parton correlators into the hadronic tensor expression in Eqs. (\ref{f:W0munu})-(\ref{f:W1Lmunu}),
by carrying out the traces we can obtain the results for the hadronic tensor up to twist-3.
The relevant traces we need are
\begin{align}
p^+ &{\rm Tr} \left[ \gamma_\alpha \hat h_{\gamma\gamma,\mu\nu}^{(0)} \right] = -4 \varrho_{\mu\nu\alpha}, \\
p^+ &{\rm Tr} \left[ \gamma_\alpha \hat h_{ZZ,\mu\nu}^{(0)} \right] = -4 c_1^q \varrho_{\mu\nu\alpha} - 4ic_3^q \varepsilon_{\alpha n \mu\nu} ,\\
p^+ &{\rm Tr} \left[ \gamma_\alpha \hat h_{\gamma Z,\mu\nu}^{(0)} \right] = -4c_V^q \varrho_{\mu\nu\alpha} - 4ic_A^q \varepsilon_{\alpha n \mu\nu}, \\
p^+ &{\rm Tr} \left[ \gamma_\alpha \gamma_5 \hat h_{\gamma\gamma,\mu\nu}^{(0)} \right] = 4 i \varepsilon_{n \alpha \mu\nu}, \\
p^+ &{\rm Tr} \left[ \gamma_\alpha \gamma_5 \hat h_{ZZ,\mu\nu}^{(0)} \right] = -4c_3^q \varrho_{\mu\nu\alpha} + 4ic_1^q \varepsilon_{n \alpha \mu\nu},\\
p^+ &{\rm Tr} \left[ \gamma_\alpha \gamma_5 \hat h_{\gamma Z,\mu\nu}^{(0)} \right] = -4c_A^q \varrho_{\mu\nu\alpha} + 4ic_V^q \varepsilon_{n \alpha \mu\nu}, \\
& {\rm Tr} \left[ \slashed{\bar n} \hat h_{\gamma\gamma,\mu\nu}^{(1)\rho} \right] = -8g_{\perp\nu}^\rho \bar n_\mu, \\
& {\rm Tr} \left[ \slashed{\bar n} \hat h_{ZZ,\mu\nu}^{(1)\rho} \right] = -8c_1^q g_{\perp\nu}^\rho \bar n_\mu - 8ic_3^q \varepsilon_{\perp\nu}^\rho \bar n_\mu,\\
& {\rm Tr} \left[ \slashed{\bar n} \hat h_{\gamma Z,\mu\nu}^{(1)\rho} \right] = -8 c_V^q g_{\perp\nu}^\rho \bar n_\mu - 8ic_A^q \varepsilon_{\perp\nu}^\rho \bar n_\mu, \\
& {\rm Tr} \left[ \slashed{\bar n} \gamma_5  \hat h_{\gamma\gamma,\mu\nu}^{(1)\rho} \right] = -8i \varepsilon_{\perp\nu}^{\rho} \bar n_\mu, \\
& {\rm Tr} \left[ \slashed{\bar n} \gamma_5 \hat h_{ZZ,\mu\nu}^{(1)\rho} \right] = -8c_3^q g_{\perp\nu}^\rho \bar n_\mu - 8i c_1^q \varepsilon_{\perp\nu}^{\rho} \bar n_\mu,\\
& {\rm Tr} \left[ \slashed{\bar n} \gamma_5 \hat h_{\gamma Z,\mu\nu}^{(1)\rho} \right] = -8c_A^q g_{\perp\nu}^\rho \bar n_\mu - 8i c_V^q \varepsilon_{\perp\nu}^{\rho} \bar n_\mu.
\end{align}
Here the $\mu\nu$-symmetric tensor  $\varrho_{\mu\nu\alpha} \equiv g_{\mu\nu} n_\alpha -  g_{\nu\alpha} n_\mu - g_{\mu\alpha} n_\nu$.
It is noted that, we can get the pure EM terms by replacing $\{ c_1^q,~ c_3^q \} \to \{1,~0\}$, and the interference terms by replacing $\{ c_1^q,~ c_3^q \} \to \{c_A^q,~c_V^q\}$ from the expressions of the pure weak interaction terms.
Therefore, we only give the pure weak interaction part $\tilde W_{ZZ}^{\mu\nu}$ up to twist-3 level for simplicity.
$\tilde W_{\gamma\gamma}^{\mu\nu}$ and $\tilde W_{\gamma Z}^{\mu\nu}$ can be obtained by the above mentioned replacement.
We first present the hadronic tensor at the leading twist (twist-2) level for completeness.
It only comes from the quark-quark correlator in Eqs.~(\ref{eq:Xi0Peven}) and (\ref{eq:Xi0Podd}).
The result is,
\begin{align}
\tilde W_{t2}^{(0)\mu\nu} =&
-\left( c_1^q g_{\perp}^{\mu\nu} + ic_3^q \varepsilon_{\perp}^{\mu\nu} \right) \Bigl(f_1+S_{LL}f_{1LL}-\frac{k_\perp \cdot \tilde S_T}{M}f^\perp_{1T}
 \nonumber\\
& \qquad\qquad\qquad+\frac{k_\perp \cdot S_{LT}}{M}f_{1LT}^{\perp}+\frac{S_{TT}^{kk}}{M^2}f_{1TT}^{\perp} \Bigr) \nonumber\\
& - \left( c_3^q g_{\perp}^{\mu\nu} + ic_1^q \varepsilon_{\perp}^{\mu\nu} \right) \Bigl(-\lambda_hg_{1L}+\frac{k_\perp\cdot S_T}{M}g^\perp_{1T} \nonumber\\
&\qquad\qquad+ \frac{k_\perp \cdot \tilde S_{LT}}{M} g_{1LT}^\perp - \frac{\tilde S_{TT}^{k k}}{M^2} g_{1TT}^\perp\Bigr). \label{f:Wt2leadingmunu}
\end{align}

The twist-3 hadronic tensor comes also from the quark-gluon-quark correlator in Eqs. (\ref{eq:Xi1Peven})-(\ref{eq:Xi1Podd}).
After using the equation of motion in Eq. (\ref{f:formulaEOM}), we get the complete hadronic tensor at twist-3 level,
\begin{widetext}
\begin{align}
(p \cdot q) \tilde W_{t3}^{\mu\nu} 
&= \Bigl[ c_1^q k_\perp^{\{\mu} \bar q^{\nu\}} + ic_3^q \tilde k_\perp^{[\mu} \bar q^{\nu]} \Bigr] \left( f^\perp + S_{LL} f_{LL}^\perp \right)- \Bigl[ c_1^q \tilde k_\perp^{\{\mu} \bar q^{\nu\}} - ic_3^q k_\perp^{[\mu} \bar q^{\nu]} \Bigr] \lambda_h f_L^\perp - \Bigl[ c_1^q \tilde S_T^{\{\mu} \bar q^{\nu\}} - ic_3^q S_T^{[\mu} \bar q^{\nu]} \Bigr] M f_T \nonumber\\
&+ \Bigl[ c_1^q S_{LT}^{\{\mu} \bar q^{\nu\}} + ic_3^q \tilde S_{LT}^{[\mu} \bar q^{\nu]} \Bigr] M f_{LT} + \Bigl[ c_1^q S_{TT}^{k\{\mu} \bar q^{\nu\}} + ic_3^q \tilde S_{TT}^{k[\mu} \bar q^{\nu]} \Bigr] f_{TT} \nonumber\\
&- \Biggl[ c_1^q \left( \frac{k_\perp\cdot \tilde S_T}{M} k_\perp^{\{\mu} \bar q^{\nu\}} - \frac{k_\perp^2}{2M} \tilde S_T^{\{\mu} \bar q^{\nu\}} \right) + ic_3^q \left(\frac{k_\perp\cdot S_T}{M} k_\perp^{[\mu} \bar q^{\nu]} - \frac{k_\perp^2}{2M}S_T^{[\mu} \bar q^{\nu]} \right) \Biggr] f_T^\perp \nonumber\\
&- \Biggl[ c_1^q \left( \frac{k_\perp\cdot S_{LT}}{M} k_\perp^{\{\mu} \bar q^{\nu\}}- \frac{k_\perp^2}{2M} S_{LT}^{\{\mu} \bar q^{\nu\}} \right) + ic_3^q \left(\frac{k_\perp\cdot S_{LT}}{M}\tilde k_\perp^{[\mu} \bar q^{\nu]} - \frac{k_\perp^2}{2M}\tilde S_{LT}^{[\mu} \bar q^{\nu]} \right) \Biggr] f_{LT}^\perp \nonumber\\
&- \Biggl[ c_1^q \left( \frac{S_{TT}^{kk}}{M^2} k_\perp^{\{\mu} \bar q^{\nu\}} - \frac{k_\perp^2}{2M^2} S_{TT}^{k\{\mu} \bar q^{\nu\}} \right) + ic_3^q \left(\frac{S_{TT}^{kk}}{M^2}\tilde k_\perp^{[\mu} \bar q^{\nu]} - \frac{k_\perp^2}{2M^2}\tilde S_{TT}^{k[\mu} \bar q^{\nu]} \right) \Biggr] f_{TT}^\perp \nonumber\\
&- \Bigl[ c_3^q \tilde k_\perp^{\{\mu} \bar q^{\nu\}} - ic_1^q k_\perp^{[\mu} \bar q^{\nu]}\Bigr] \left( g^\perp + S_{LL} g_{LL}^\perp \right)- \Bigl[ c_3^q k_\perp^{\{\mu} \bar q^{\nu\}} + ic_1^q \tilde k_\perp^{[\mu} \bar q^{\nu]}\Bigr] \lambda_h g_L^\perp - \Bigl[ c_3^q S_T^{\{\mu} \bar q^{\nu\}} + ic_1^q \tilde S_T^{[\mu} \bar q^{\nu]} \Bigr] M g_T \nonumber\\
&- \Bigl[ c_3^q \tilde S_{LT}^{\{\mu} \bar q^{\nu\}} - ic_1^q S_{LT}^{[\mu} \bar q^{\nu]} \Bigr] M g_{LT} - \Bigl[ c_3^q \tilde S_{TT}^{k\{\mu} \bar q^{\nu\}} - ic_1^q S_{TT}^{k[\mu} \bar q^{\nu]} \Bigr] g_{TT} \nonumber\\
&+ \Biggl[ c_3^q \left( \frac{k_\perp\cdot S_T}{M} k_\perp^{\{\mu} \bar q^{\nu\}} - \frac{k_\perp^2}{2M} S_T^{\{\mu} \bar q^{\nu\}} \right) + ic_1^q \left(\frac{k_\perp\cdot S_T}{M} \tilde k_\perp^{[\mu} \bar q^{\nu]} - \frac{k_\perp^2}{2M} \tilde S_T^{[\mu} \bar q^{\nu]} \right) \Biggr] g_T^\perp \nonumber\\
&- \Biggl[ c_3^q \left( \frac{k_\perp\cdot \tilde S_{LT}}{M} k_\perp^{\{\mu} \bar q^{\nu\}} - \frac{k_\perp^2}{2M} \tilde S_{LT}^{\{\mu} \bar q^{\nu\}} \right) + ic_1^q \left(\frac{k_\perp\cdot S_{LT}}{M} k_\perp^{[\mu} \bar q^{\nu]} - \frac{k_\perp^2}{2M} S_{LT}^{[\mu} \bar q^{\nu]} \right) \Biggr] g_{LT}^\perp \nonumber\\
&- \Biggl[ c_3^q \left( \frac{k_\perp\cdot \tilde S_{TT}^{k}}{M^2} k_\perp^{\{\mu} \bar q^{\nu\}}- \frac{k_\perp^2}{2M^2} \tilde S_{TT}^{k\{\mu} \bar q^{\nu\}} \right) + ic_1^q \left(\frac{k_\perp\cdot S_{TT}^{k}}{M^2} k_\perp^{[\mu} \bar q^{\nu]} - \frac{k_\perp^2}{2M^2} S_{TT}^{k[\mu} \bar q^{\nu]} \right) \Biggr] g_{TT}^\perp, \label{f:Wt3munu}
\end{align}
where $\bar q^\mu = q^\mu + 2xp^\mu$.
From $q\cdot\bar q = q\cdot k_\perp = 0$ and $q\cdot S_T = q\cdot S_{LT} = q\cdot S_{TT}^{k}/M = 0$, we see clearly that the full twist-3 hadronic tensor satisfies current conservation, $q_\mu \tilde W^{\mu\nu}_{t3} = q_\nu \tilde W^{\mu\nu}_{t3} = 0$.

\section{The cross section up to twist-3} \label{sec:crosssection}

Substituting the leading twist hadronic tensor in Eq.~(\ref{f:Wt2leadingmunu}) and the leptonic tensor into Eq.~(\ref{f:crosssection}) yields the leading twist cross section.
Here, we also give the expressions explicitly for the weak interaction part
\begin{align}
  &\frac{d\sigma_{t2}^{ZZ}}{dx dy d\psi d^2 k_\perp^\prime} =\frac{\alpha_{\rm em}^2 \chi}{y Q^2}\Biggl\{ \left[T_0^q(y) -\lambda_e \tilde T_0^q(y) \right] (f_1+S_{LL}f_{1LL})- \left[\tilde T_1^q(y) -\lambda_e T_1^q(y) \right] \lambda_h g_{1L} \nonumber\\
  &+|S_T|k_{\perp M}\Big[\sin(\varphi-\varphi_S) \bigl(T_0^q(y) -\lambda_e \tilde T_0^q(y) \bigr) f^\perp_{1T}-\cos(\varphi-\varphi_S) \bigl(\tilde T_1^q(y) -\lambda_e T_1^q(y) \bigr)g^\perp_{1T}\Big] \nonumber\\
  &-|S_{LT}|k_{\perp M}\Big[\sin(\varphi-\varphi_{LT}) \bigl(\tilde T_1^q(y) -\lambda_e T_1^q(y) \bigr) g^\perp_{1LT}+\cos(\varphi-\varphi_{LT}) \bigl(T_0^q(y) -\lambda_e \tilde T_0^q(y) \bigr)f^\perp_{1LT}\Big] \nonumber\\
  &-|S_{TT}|k_{\perp M}^2\Big[\sin(2\varphi-2\varphi_{TT}) \bigl(\tilde T_1^q(y) -\lambda_e T_1^q(y) \bigr) g^\perp_{1TT}-\cos(2\varphi-2\varphi_{TT}) \bigl(T_0^q(y) -\lambda_e \tilde T_0^q(y) \bigr)f^\perp_{1TT}\Big]
 \Biggr\},
\end{align}
where we have defined $k_{\perp M} = |\vec k_\perp|/M$, and
\begin{align}
  & T_0^q(y) = c_1^e c_1^q A(y) + c_3^e c_3^q C(y), \nonumber\\
  & \tilde T_0^q(y) = c_3^e c_1^q A(y) + c_1^e c_3^q C(y), \nonumber\\
  & T_1^q(y) = c_3^e c_3^q A(y) + c_1^e c_1^q C(y), \nonumber\\
  & \tilde T_1^q(y) = c_1^e c_3^q A(y) + c_3^e c_1^q C(y),
\label{eq:T0T1}
\end{align}
to simplify the expressions.
$T_i^q(y)$'s and $\tilde T_i^q(y)$'s are related to space reflection even and odd structure respectively in the cross section.
For EM interaction, it requires $c_3^{e/q} = 0$ and $c_1^{e/q} = 1$. In this case, only $T_0^q(y)$ and $T_1^q(y)$ are left, and $T_0^q(y)=A(y)$, $T_1^q(y)=C(y)$. For the interference terms, we need to set $c_3^{e/q}=c_A^{e/q}$ and $c_1^{e/q}=c_V^{e/q}$.
The kinematic factors are also different.
To make it transparent, we can get the EM and interference cross sections by replacing the parameters in the weak interaction cross section according to Tab.~\ref{tab:replacing}
\begin{table}
\renewcommand\arraystretch{1.5}
\begin{tabular}{c|c|c|c}
\hline
~~Interactions ~~  & $A_r$   & $L^{\mu\nu}_r$   & $W^{\mu\nu}_r$  \\ \hline \hline
$ZZ$ & $\chi$ & $c_1^e,~c_3^e$ & $c_1^q,~c_3^q$  \\ \hline
$\gamma Z$  & ~~$\chi\to \chi_{int}$ ~~& ~~$c_1^e\to c_V^e,~c_3^e\to c_A^e$~~ &~~ $c_1^q\to c_V^q,~c_3^q\to c_A^q$~~ \\ \hline
$\gamma\gamma$ & $\chi\to e_q^2$  & $c_1^e\to 1,~c_3^e\to 0$  & $c_1^q\to 1,~c_3^q\to 0$ \\ \hline
\end{tabular}
\caption{Relations of kinematic factors between weak, EM and interference interactions.}
\label{tab:replacing}
\end{table}

Similarly, substituting the twist-3 hadronic tensor in Eq. (\ref{f:Wt3munu}) and the leptonic tensor into Eq. (\ref{f:crosssection}) yields the twist-3 cross section. It is given by
\begin{align}
  \frac{d\sigma_{t3}^{ZZ}}{dx dy d\psi d^2 k_\perp^\prime} =& -\frac{\alpha_{\rm{em}}^2 \chi }{y Q^2}2x\kappa_M \Biggl\{ k_{\perp M}\cos\varphi \bigl(T_2^q(y)-\lambda_e \tilde T_2^q(y)\bigr)(f^\perp+S_{LL}f^\perp_{LL})+k_{\perp M}\sin\varphi \bigl(\tilde T_3^q(y)-\lambda_e T_3^q(y)\bigr)(g^\perp+S_{LL}g^\perp_{LL}) \nonumber\\
  &+\lambda_h k_{\perp M}\Big[\sin\varphi \bigl( T_2^q(y)-\lambda_e \tilde T_2^q(y)\bigr)f^\perp_L - \cos\varphi \bigl( \tilde T_3^q(y)-\lambda_e T_3^q(y)\bigr)g_L^\perp\Big] \nonumber\\
  &+|S_T|\Big[\sin\varphi_S \bigl(  T_2^q(y)-\lambda_e \tilde T_2^q(y)\bigr)f_T -\cos\varphi_S \bigl( \tilde T_3^q(y)-\lambda_e T_3^q(y)\bigr)g_T  \nonumber\\
  &\quad +\sin(2\varphi-\varphi_S) \bigl( T_2^q(y)-\lambda_e \tilde T_2^q(y)\bigr)\frac{k_{\perp M}^2}{2}f^\perp_T -\cos(2\varphi-\varphi_S) \bigl( \tilde T_3^q(y)-\lambda_e T_3^q(y)\bigr)\frac{k_{\perp M}^2}{2}g^\perp_T \Big] \nonumber\\
  &+|S_{LT}|\Big[\sin\varphi_{LT} \bigl( \tilde T_3^q(y)-\lambda_e T_3^q(y)\bigr)g_{LT} +\cos\varphi_{LT} \bigl( T_2^q(y)-\lambda_e \tilde T_2^q(y)\bigr)f_{LT}  \nonumber\\
  &\quad +\sin(2\varphi-\varphi_{LT}) \bigl( \tilde T_3^q(y)-\lambda_e T_3^q(y)\bigr)\frac{k_{\perp M}^2}{2} g^\perp_{LT} + \cos(2\varphi-\varphi_{LT}) \bigl( T_2^q(y)-\lambda_e \tilde T_2^q(y)\bigr)\frac{k_{\perp M}^2}{2} f^\perp_{LT} \Big] \nonumber\\
  &+|S_{TT}|\Big[\sin(\varphi-2\varphi_{TT}) \bigl( \tilde T_3^q(y)-\lambda_e T_3^q(y)\bigr)k_{\perp M} g_{TT} -\cos(\varphi-2\varphi_{TT}) \bigl( T_2^q(y)-\lambda_e \tilde T_2^q(y)\bigr)k_{\perp M} f_{TT} \nonumber\\
  &\quad -\sin(3\varphi-2\varphi_{TT}) \bigl( \tilde T_3^q(y)-\lambda_e T_3^q(y)\bigr)\frac{k_{\perp M}^3}{2} g^\perp_{TT} -\cos(3\varphi-2\varphi_{TT}) \bigl( T_2^q(y)-\lambda_e \tilde T_2^q(y)\bigr)\frac{k_{\perp M}^3}{2} f^\perp_{TT} \Big]
 \Biggr\},
\end{align}
\end{widetext}
where we have defined $\kappa_M=M/Q$ to simplify the expression.
We have also defined
\begin{align}
  & T_2^q(y) = c_1^e c_1^q B(y) + c_3^e c_3^q D(y), \nonumber\\
  & \tilde T_2^q(y) = c_3^e c_1^q B(y) + c_1^e c_3^q D(y), \nonumber\\
  & T_3^q(y) = c_3^e c_3^q B(y) + c_1^e c_1^q D(y), \nonumber\\
  & \tilde T_3^q(y) = c_1^e c_3^q B(y) + c_3^e c_1^q D(y).
\label{eq:T2T3}
\end{align}

It is also straightforward to obtain the interference and EM contributions by doing the corresponding replacements.
To further unify the notations, we define $T_{i,r}^q(y)$'s and $\tilde T_{i,r}^q(y)$'s with $r=ZZ$, $\gamma Z$ and $\gamma\gamma$. For the weak interaction, we have $T_{i,ZZ}^q(y)$'s and $\tilde T_{i,ZZ}^q(y)$'s defined as $T_{i}^q(y)$'s and $\tilde T_{i}^q(y)$'s given in Eqs.~(\ref{eq:T0T1}) and (\ref{eq:T2T3}) respectively.
For $\gamma Z$ and $\gamma\gamma$ parts, according to Tab.~\ref{tab:replacing}, we have:
\begin{align}
  & T_{0,\gamma Z}^q(y) = c_V^e c_V^q A(y) + c_A^e c_A^q C(y), \nonumber\\
  & \tilde T_{0,\gamma Z}^q(y) = c_A^e c_V^q A(y) + c_V^e c_A^q C(y), \nonumber\\
  & T_{1,\gamma Z}^q(y) = c_A^e c_A^q A(y) + c_V^e c_V^q C(y), \nonumber\\
  & \tilde T_{1,\gamma Z}^q(y) = c_V^e c_A^q A(y) + c_A^e c_V^q C(y), \nonumber\\
  & T_{2,\gamma Z}^q(y) = c_V^e c_V^q B(y) + c_A^e c_A^q D(y), \nonumber\\
  & \tilde T_{2,\gamma Z}^q(y) = c_A^e c_V^q B(y) + c_V^e c_A^q D(y), \nonumber\\
  & T_{3,\gamma Z}^q(y) = c_A^e c_A^q B(y) + c_V^e c_V^q D(y), \nonumber\\
  & \tilde T_{3,\gamma Z}^q(y) = c_V^e c_A^q B(y) + c_A^e c_V^q D(y),
\label{eq:Ts-gammaZ}
\end{align}
and
\begin{align}
  & T_{0,\gamma\gamma}^q(y) = A(y), \quad \tilde T_{0,\gamma\gamma}^q(y) = 0 \nonumber\\
  & T_{1,\gamma\gamma}^q(y) = C(y), \quad \tilde T_{1,\gamma\gamma}^q(y) = 0 \nonumber\\
  & T_{2,\gamma\gamma}^q(y) = B(y), \quad \tilde T_{2,\gamma\gamma}^q(y) = 0 \nonumber\\
  & T_{3,\gamma\gamma}^q(y) = D(y), \quad \tilde T_{3,\gamma\gamma}^q(y) = 0.
\label{eq:Ts-gammagamma}
\end{align}
We see that only half of the terms will survive if only EM interaction is considered.

\section{Structure functions and azimuthal asymmetries results up to twist-3} \label{sec:result}

In Sec. \ref{sec:formalism}, we have presented the general form of the cross section in terms of structure functions.
In the previous section we have also presented the cross section in terms of the gauge invariant TMD PDFs.
They match to each other.
In this section, we present the structure functions and azimuthal asymmetries results in terms of the TMD PDFs.

\subsection{Structure functions results}

We first present the structure functions in terms of gauge invariant PDFs. For the leading twist part, we have,
\begin{align}
  & W^T_{U,U} = c_1^e c_1^q f_1, \\
  & W_{U,U} = c_3^e c_3^q f_1, \\
  & \tilde W^T_{L,U} = - c_3^e c_1^q f_1, \\
  & \tilde W_{L,U} = - c_1^e c_3^q f_1, \\
  & \tilde W^T_{U,L} = - c_1^e c_3^q g_{1L}, \\
  & \tilde W_{U,L} = - c_3^e c_1^q g_{1L}, \\
  & W^T_{L,L} = c_3^e c_3^q g_{1L}, \\
  & W_{L,L} = c_1^e c_1^q g_{1L}, \\
  & W^T_{U,LL} = c_1^e c_1^q f_{1LL}, \\
  & W_{U,LL} = c_3^e c_3^q f_{1LL}, \\
  & \tilde W^T_{L,LL} = - c_3^e c_1^q f_{1LL}, \\
  & \tilde W_{L,LL} = - c_1^e c_3^q f_{1LL}, \\
  & \tilde W_{U,T}^{T,\cos(\varphi-\varphi_S)} = - c_1^e c_3^q k_{\perp M} g^\perp_{1T}, \\
  & \tilde W_{U,T}^{\cos(\varphi-\varphi_S)} = - c_3^e c_1^q k_{\perp M} g^\perp_{1T}, \\
  & W_{U,T}^{T,\sin(\varphi-\varphi_S)} = c_1^e c_1^q k_{\perp M} f^\perp_{1T}, \\
  & W_{U,T}^{\sin(\varphi-\varphi_S)} = c_3^e c_3^q k_{\perp M} f^\perp_{1T}, \\
  & W_{L,T}^{T,\cos(\varphi-\varphi_S)} =  c_3^e c_3^q k_{\perp M} g^\perp_{1T}, \\
  & W_{L,T}^{\cos(\varphi-\varphi_S)} =  c_1^e c_1^q k_{\perp M} g^\perp_{1T}, \\
  & \tilde W_{L,T}^{T,\sin(\varphi-\varphi_S)} = -c_3^e c_1^q k_{\perp M} f^\perp_{1T}, \\
  & \tilde W_{L,T}^{\sin(\varphi-\varphi_S)} = -c_1^e c_3^q k_{\perp M} f^\perp_{1T}, \\
  & W_{U,LT}^{T,\cos(\varphi-\varphi_{LT})} = - c_1^e c_1^q k_{\perp M} f^\perp_{1LT}, \\
  & W_{U,LT}^{\cos(\varphi-\varphi_{LT})} = - c_3^e c_3^q k_{\perp M} f^\perp_{1LT}, \\
  & \tilde W_{U,LT}^{T,\sin(\varphi-\varphi_{LT})} = -c_1^e c_3^q k_{\perp M} g^\perp_{1LT}, \\
  & \tilde W_{U,LT}^{\sin(\varphi-\varphi_{LT})} = -c_3^e c_1^q k_{\perp M} g^\perp_{1LT}, \\
  & \tilde W_{L,LT}^{T,\cos(\varphi-\varphi_{LT})} =  c_3^e c_1^q k_{\perp M} f^\perp_{1LT}, \\
  & \tilde W_{L,LT}^{\cos(\varphi-\varphi_{LT})} =  c_1^e c_3^q k_{\perp M} f^\perp_{1LT}, \\
  & W_{L,LT}^{T,\sin(\varphi-\varphi_{LT})} = c_3^e c_3^q k_{\perp M} g^\perp_{1LT}, \\
  & W_{L,LT}^{\sin(\varphi-\varphi_{LT})} = c_1^e c_1^q k_{\perp M} g^\perp_{1LT}, \\
  & W_{U,TT}^{T,\cos(2\varphi-2\varphi_{TT})} =  c_1^e c_1^q k^2_{\perp M} f^\perp_{1TT}, \\
  & W_{U,TT}^{\cos(2\varphi-2\varphi_{TT})} =  c_3^e c_3^q k^2_{\perp M} f^\perp_{1TT}, \\
  & \tilde W_{U,TT}^{T,\sin(2\varphi-2\varphi_{TT})} = -c_1^e c_3^q k^2_{\perp M} g^\perp_{1TT}, \\
  & \tilde W_{U,TT}^{\sin(2\varphi-2\varphi_{TT})} = -c_3^e c_1^q k^2_{\perp M} g^\perp_{1TT}, \\
  & \tilde W_{L,TT}^{T,\cos(2\varphi-2\varphi_{TT})} =  -c_3^e c_1^q k^2_{\perp M} f^\perp_{1TT}, \\
  & \tilde W_{L,TT}^{\cos(2\varphi-2\varphi_{TT})} =  -c_1^e c_3^q k^2_{\perp M} f^\perp_{1TT}, \\
  & W_{L,TT}^{T,\sin(2\varphi-2\varphi_{TT})} = c_3^e c_3^q k^2_{\perp M} g^\perp_{1TT}, \\
  & W_{L,TT}^{\sin(2\varphi-2\varphi_{TT})} = c_1^e c_1^q k_{\perp M} g^\perp_{1TT}.
\end{align}
In total we have 36 structure functions which contribute to the leading twist.
If only the EM interaction is taken into account, only one fourth of them which are related to $c_1^e c_1^q = 1$ are left.

For the twist-3 part, we have
\begin{align}
  & W_{U,U1}^{\cos\varphi} = -2x\kappa_M k_{\perp M} c_1^ec_1^q f^\perp, \\
  & \tilde W_{U,U1}^{\sin\varphi} = -2x\kappa_M k_{\perp M} c_1^ec_3^q g^\perp, \\
  & W_{U,U2}^{\cos\varphi} = -2x\kappa_M k_{\perp M} c_3^ec_3^q f^\perp, \\
  & \tilde W_{U,U2}^{\sin\varphi} = -2x\kappa_M k_{\perp M} c_3^ec_1^q g^\perp, \\
  & \tilde W_{L,U1}^{\cos\varphi} = 2x\kappa_M k_{\perp M} c_3^ec_1^q f^\perp, \\
  & W_{L,U1}^{\sin\varphi} = 2x\kappa_M k_{\perp M} c_3^ec_3^q g^\perp, \\
  & \tilde W_{L,U2}^{\cos\varphi} = 2x\kappa_M k_{\perp M} c_1^ec_3^q f^\perp, \\
  & W_{L,U2}^{\sin\varphi} = 2x\kappa_M k_{\perp M} c_1^ec_1^q g^\perp, \\
  & \tilde W_{U,L1}^{\cos\varphi} = 2x\kappa_M k_{\perp M} c_1^ec_3^q g_L^\perp, \\
  & W_{U,L1}^{\sin\varphi} = -2x\kappa_M k_{\perp M} c_1^ec_1^q f_L^\perp, \\
  & \tilde W_{U,L2}^{\cos\varphi} = 2x\kappa_M k_{\perp M} c_3^ec_1^q g_L^\perp, \\
  & W_{U,L2}^{\sin\varphi} = -2x\kappa_M k_{\perp M} c_3^ec_3^q f_L^\perp, \\
  & W_{L,L1}^{\cos\varphi} = -2x\kappa_M k_{\perp M} c_3^ec_3^q g_L^\perp, \\
  & \tilde W_{L,L1}^{\sin\varphi} = 2x\kappa_M k_{\perp M} c_3^ec_1^q f_L^\perp, \\
  & W_{L,L2}^{\cos\varphi} = -2x\kappa_M k_{\perp M} c_1^ec_1^q g_L^\perp, \\
  & \tilde W_{L,L2}^{\sin\varphi} = 2x\kappa_M k_{\perp M} c_1^ec_3^q f_L^\perp, \\
  & W_{U,LL1}^{\cos\varphi} = -2x\kappa_M k_{\perp M} c_1^ec_1^q f_{LL}^\perp, \\
  & \tilde W_{U,LL1}^{\sin\varphi} = -2x\kappa_M k_{\perp M} c_1^ec_3^q g_{LL}^\perp, \\
  & W_{U,LL2}^{\cos\varphi} = -2x\kappa_M k_{\perp M} c_3^ec_3^q f_{LL}^\perp, \\
  & \tilde W_{U,LL2}^{\sin\varphi} = -2x\kappa_M k_{\perp M} c_3^ec_1^q g_{LL}^\perp, \\
  & \tilde W_{L,LL1}^{\cos\varphi} = 2x\kappa_M k_{\perp M} c_3^ec_1^q f_{LL}^\perp, \\
  & W_{L,LL1}^{\sin\varphi} = 2x\kappa_M k_{\perp M} c_3^ec_3^q g_{LL}^\perp, \\
  & \tilde W_{L,LL2}^{\cos\varphi} = 2x\kappa_M k_{\perp M} c_1^ec_3^q f_{LL}^\perp, \\
  & W_{L,LL2}^{\sin\varphi} = 2x\kappa_M k_{\perp M} c_1^ec_1^q g_{LL}^\perp, \\
  & \tilde W_{U,T1}^{\cos\varphi_S} = 2x\kappa_M  c_1^ec_3^q g_T, \\
  & W_{U,T1}^{\sin\varphi_S} = -2x\kappa_M  c_1^ec_1^q f_T, \\
  & \tilde W_{U,T2}^{\cos\varphi_S} = 2x\kappa_M  c_3^ec_1^q g_T, \\
  & W_{U,T2}^{\sin\varphi_S} = -2x\kappa_M  c_3^ec_3^q f_T, \\
  & \tilde W_{U,T1}^{\cos(2\varphi-\varphi_S)} = x\kappa_M k^2_{\perp M} c_1^ec_3^q g_T^\perp, \\
  & W_{U,T1}^{\sin(2\varphi-\varphi_S)} = -x\kappa_M k^2_{\perp M} c_1^ec_1^q f_T^\perp, \\
  & \tilde W_{U,T2}^{\cos(2\varphi-\varphi_S)} = x\kappa_M k^2_{\perp M} c_3^ec_1^q g_T^\perp, \\
  & W_{U,T2}^{\sin(2\varphi-\varphi_S)} = -x\kappa_M k^2_{\perp M} c_3^ec_3^q f_T^\perp, \\
  & W_{L,T1}^{\cos\varphi_S} = -2x\kappa_M  c_3^ec_3^q g_T, \\
  & \tilde W_{L,T1}^{\sin\varphi_S} = 2x\kappa_M  c_3^ec_1^q f_T, \\
  & W_{L,T2}^{\cos\varphi_S} = -2x\kappa_M  c_1^ec_1^q g_T, \\
  & \tilde W_{L,T2}^{\sin\varphi_S} = 2x\kappa_M  c_1^ec_3^q f_T, \\
  & W_{L,T1}^{\cos(2\varphi-\varphi_S)} = -x\kappa_M k^2_{\perp M} c_3^ec_3^q g_T^\perp, \\
  & \tilde W_{L,T1}^{\sin(2\varphi-\varphi_S)} = x\kappa_M k^2_{\perp M} c_3^ec_1^q f_T^\perp, \\
  & W_{L,T2}^{\cos(2\varphi-\varphi_S)} = -x\kappa_M k^2_{\perp M} c_1^ec_1^q g_T^\perp, \\
  & \tilde W_{L,T2}^{\sin(2\varphi-\varphi_S)} = x\kappa_M k^2_{\perp M} c_1^ec_3^q f_T^\perp, \\
  & W_{U,LT1}^{\cos\varphi_{LT}} = -2x\kappa_M  c_1^ec_1^q f_{LT}, \\
  & \tilde W_{U,LT1}^{\sin\varphi_{LT}} = -2x\kappa_M  c_1^ec_3^q g_{LT}, \\
  & W_{U,LT2}^{\cos\varphi_{LT}} = -2x\kappa_M  c_3^ec_3^q f_{LT}, \\
  & \tilde W_{U,LT2}^{\sin\varphi_{LT}} = -2x\kappa_M  c_3^ec_1^q g_{LT}, \\
  & W_{U,LT1}^{\cos(2\varphi-\varphi_{LT})} = -x\kappa_M k^2_{\perp M} c_1^ec_1^q f_{LT}^\perp, \\
  & \tilde W_{U,LT1}^{\sin(2\varphi-\varphi_{LT})} = -x\kappa_M k^2_{\perp M} c_1^ec_3^q g_{LT}^\perp, \\
  & W_{U,LT2}^{\cos(2\varphi-\varphi_{LT})} = -x\kappa_M k^2_{\perp M} c_3^ec_3^q f_{LT}^\perp, \\
  & \tilde W_{U,LT2}^{\sin(2\varphi-\varphi_{LT})} = -x\kappa_M k^2_{\perp M} c_3^ec_1^q g_{LT}^\perp, \\
  & \tilde W_{L,LT1}^{\cos\varphi_{LT}} = 2x\kappa_M  c_3^ec_1^q f_{LT}, \\
  & W_{L,LT1}^{\sin\varphi_{LT}} = 2x\kappa_M  c_3^ec_3^q g_{LT}, \\
  & \tilde W_{L,LT2}^{\cos\varphi_{LT}} = 2x\kappa_M  c_1^ec_3^q f_{LT}, \\
  & W_{L,LT2}^{\sin\varphi_{LT}} = -2x\kappa_M  c_1^ec_1^q g_{LT}, \\
  & \tilde W_{L,LT1}^{\cos(2\varphi-\varphi_{LT})} = x\kappa_M k^2_{\perp M} c_3^ec_1^q f_{LT}^\perp, \\
  & W_{L,LT1}^{\sin(2\varphi-\varphi_{LT})} = x\kappa_M k^2_{\perp M} c_3^ec_3^q g_{LT}^\perp, \\
  & \tilde W_{L,LT2}^{\cos(2\varphi-\varphi_{LT})} = x\kappa_M k^2_{\perp M} c_1^ec_3^q f_{LT}^\perp, \\
  & W_{L,LT2}^{\sin(2\varphi-\varphi_{LT})} = x\kappa_M k^2_{\perp M} c_1^ec_1^q g_{LT}^\perp, \\
& W_{U,TT1}^{\cos(\varphi-2\varphi_{TT})} = 2x\kappa_M k_{\perp M} c_1^e c_1^q f_{TT}, \\
& W_{U,TT2}^{\cos(\varphi-2\varphi_{TT})} = 2x\kappa_M k_{\perp M} c_3^e c_3^q f_{TT}, \\
& \tilde W_{U,TT1}^{\sin(\varphi-2\varphi_{TT})} = -2x\kappa_M k_{\perp M} c_1^e c_3^q g_{TT}, \\
& \tilde W_{U,TT2}^{\sin(\varphi-2\varphi_{TT})} = -2x\kappa_M k_{\perp M} c_3^e c_1^q g_{TT}, \\
& W_{U,TT1}^{\cos(3\varphi-2\varphi_{TT})} = x\kappa_M k_{\perp M}^3 c_1^e c_1^q f_{TT}^\perp, \\
& W_{U,TT2}^{\cos(3\varphi-2\varphi_{TT})} = x\kappa_M k_{\perp M}^3 c_3^e c_3^q f_{TT}^\perp, \\
& \tilde W_{U,TT1}^{\sin(3\varphi-2\varphi_{TT})} = x\kappa_M k_{\perp M}^3 c_1^e c_3^q g_{TT}^\perp, \\
& \tilde W_{U,TT2}^{\sin(3\varphi-2\varphi_{TT})} = x\kappa_M k_{\perp M}^3 c_3^e c_1^q g_{TT}^\perp, \\
& \tilde W_{L,TT1}^{\cos(\varphi-2\varphi_{TT})} = -2x\kappa_M k_{\perp M} c_3^e c_1^q f_{TT}, \\
& \tilde W_{L,TT2}^{\cos(\varphi-2\varphi_{TT})} = -2x\kappa_M k_{\perp M} c_1^e c_3^q f_{TT}, \\
& W_{L,TT1}^{\sin(\varphi-2\varphi_{TT})} = 2x\kappa_M k_{\perp M} c_3^e c_3^q g_{TT}, \\
& W_{L,TT2}^{\sin(\varphi-2\varphi_{TT})} = 2x\kappa_M k_{\perp M} c_1^e c_1^q g_{TT}, \\
& \tilde W_{L,TT1}^{\cos(3\varphi-2\varphi_{TT})} = -x\kappa_M k_{\perp M}^3 c_3^e c_1^q f_{TT}^\perp, \\
& \tilde W_{L,TT2}^{\cos(3\varphi-2\varphi_{TT})} = -x\kappa_M k_{\perp M}^3 c_1^e c_3^q f_{TT}^\perp, \\
& W_{L,TT1}^{\sin(3\varphi-2\varphi_{TT})} = -x\kappa_M k_{\perp M}^3 c_3^e c_3^q g_{TT}^\perp, \\
& W_{L,TT2}^{\sin(3\varphi-2\varphi_{TT})} = -x\kappa_M k_{\perp M}^3 c_1^e c_1^q g_{TT}^\perp.
\end{align}
In total we have 72 structure functions contribute at twist-3.
Also, one fourth of them are left if only EM interaction is taken into account.

One can get the full structure functions results measured in experiments by summing the weak, EM and interference terms together. To this end, we would better redefine the structure functions to include also the kinematic factor $A_r$'s. The results is simple to get, e.g.,
\begin{align}
W_{U,U1}^{\cos\varphi} = -2x\kappa_M k_{\perp M} \left( e_q^2 + c_1^ec_1^q \chi + c_V^e c_V^q \chi_{int}  \right) f^\perp.
\end{align}

\subsection{Azimuthal asymmetries from unpolarized electron beam}

In addition to structure functions, we also calculate the azimuthal asymmetries results.
We consider both the unpolarized beam ($\lambda_e=0$) and the polarized beam ($\lambda_e=\pm1$) cases.
They contribute to different azimuthal asymmetries results.
We first consider the unpolarized case.
The azimuthal asymmetry is defined as, e.g.,
\begin{align}
  \langle \sin\varphi \rangle_{U,U}=\frac{\int d\tilde\sigma \sin\varphi d\varphi}{\int d\tilde\sigma d\varphi},
\end{align}
for the unpolarized or longitudinally polarized target case, and
\begin{align}
  \langle \sin(\varphi-\varphi_S) \rangle_{U,T}=\frac{\int d\tilde\sigma \sin(\varphi-\varphi_S)d\varphi d\varphi_S}{\int d\tilde\sigma d\varphi d\varphi_S},
\end{align}
for the transversely polarized target case.
$d\tilde\sigma$ is used to denote $\frac{d\sigma}{dx dy d\psi d^2 k_\perp^\prime}$,
and $d\varphi_S\approx d\psi$
which integration corresponds to take the average over the out going electron's azimuthal angle~\cite{Diehl:2005pc,Bacchetta:2006tn}.
The subscripts such as ($U, T$) denote the polarizations of the lepton beam and the target, respectively.
At the leading twist, there are six polarization dependent azimuthal asymmetries which are given by (the sum over $r=ZZ$, $\gamma Z$ and $\gamma\gamma$ is implicit in the numerator and the denominator respectively)
\begin{align}
 & \langle \sin(\varphi-\varphi_S) \rangle_{U,T} = k_{\perp M} \frac{ A_r T_{0,r}^q(y) f^\perp_{1T}}{2A_r T_{0,r}^q(y) f_1}, \\
 & \langle \cos(\varphi-\varphi_S) \rangle_{U,T} = - k_{\perp M} \frac{A_r \tilde T_{1,r}^q(y)g^\perp_{1T}}{2A_r T_{0,r}^q(y) f_1}, \\
 & \langle \sin(\varphi-\varphi_{LT}) \rangle_{U,LT} = -k_{\perp M} \frac{A_r \tilde T_{1,r}^q(y)g^\perp_{1LT}}{2A_r T_{0,r}^q(y)f_1}, \\
 & \langle \cos(\varphi-\varphi_{LT}) \rangle_{U,LT} = - k_{\perp M} \frac{A_r T_{0,r}^q(y)f^\perp_{1LT}}{2A_r T_{0,r}^q(y)f_1}, \\
 & \langle \sin(2\varphi-\varphi_{TT}) \rangle_{U,TT} = - k_{\perp M}^2 \frac{A_r \tilde T_{1,r}^q(y)g^\perp_{1TT}}{2A_r T_{0,r}^q(y)f_1}, \\
 & \langle \cos(2\varphi-\varphi_{TT}) \rangle_{U,TT} = k_{\perp M}^2 \frac{A_r T_{0,r}^q(y)f^\perp_{1TT}}{2A_r T_{0,r}^q(y)f_1}.
\end{align}
We find that there are three parity-violating azimuthal asymmetry modulations among the six in total. If only electromagnetic interactions are considered only three parity conserved modulations are left.
At twist-3, we have 18 azimuthal asymmetries. They are given by
\begin{align}
  & \langle \cos\varphi \rangle_{U,U} = -x\kappa_M k_{\perp M} \frac{ A_r T_{2,r}^q(y)}{A_r T_{0,r}^q(y)}\frac{f^\perp}{f_1}, \\
  & \langle \sin\varphi \rangle_{U,U} = -x\kappa_M k_{\perp M} \frac{A_r \tilde T_{3,r}^q(y)}{A_r T_{0,r}^q(y)}\frac{g^\perp}{f_1}, \\
  & \langle \cos\varphi \rangle_{U,L} = -x\kappa_M k_{\perp M} \frac{A_r T_{2,r}^q(y) f^\perp -\lambda_h A_r \tilde T_{3,r}^q(y)g^\perp_L}{A_r T_{0,r}^q(y) f_1}, \\
  & \langle \sin\varphi \rangle_{U,L} = -x\kappa_M k_{\perp M} \frac{A_r \tilde T_{3,r}^q(y) g^\perp + \lambda_h A_r T_{2,r}^q(y)f^\perp_L}{A_r T_{0,r}^q(y)f_1}, \\
  & \langle \cos\varphi \rangle_{U,LL} = -x\kappa_M k_{\perp M} \frac{A_r T_{2,r}^q(y)(f^\perp + S_{LL}f^\perp_{LL})}{A_r T_{0,r}^q(y)f_1}, \\
  & \langle \sin\varphi \rangle_{U,LL} = -x\kappa_M k_{\perp M} \frac{A_r \tilde T_{3,r}^q(y) (g^\perp + S_{LL}g^\perp_{LL})}{A_r T_{0,r}^q(y)f_1}, \\
  & \langle \cos\varphi_S \rangle_{U,T} =  x\kappa_M\frac{ A_r \tilde T_{3,r}^q(y)g_T}{A_r T_{0,r}^q(y)f_1}, \\
  & \langle \sin\varphi_S \rangle_{U,T} = -x\kappa_M\frac{ A_r T_{2,r}^q(y)f_T}{A_r T_{0,r}^q(y)f_1}, \\
  & \langle \cos(2\varphi-\varphi_S) \rangle_{U,T} = x\kappa_M k_{\perp M}^2\frac{ A_r \tilde T_{3,r}^q(y)g^\perp_T}{2A_r T_{0,r}^q(y)f_1}, \\
  & \langle \sin(2\varphi-\varphi_S) \rangle_{U,T} = -x\kappa_M k_{\perp M}^2\frac{ A_r T_{2,r}^q(y)f^\perp_T}{2A_r T_{0,r}^q(y)f_1}, \\
  & \langle \cos\varphi_{LT} \rangle_{U,LT} = - x\kappa_M\frac{ A_r T_{2,r}^q(y)f_{LT}}{A_r T_{0,r}^q(y)f_1}, \\
  & \langle \sin\varphi_{LT} \rangle_{U,LT} = -x\kappa_M\frac{ A_r \tilde T_{3,r}^q(y)g_{LT}}{A_r T_{0,r}^q(y)f_1}, \\
  & \langle \cos(2\varphi-\varphi_{LT}) \rangle_{U,LT} = -x\kappa_M k_{\perp M}^2\frac{ A_r T_{2,r}^q(y)f^\perp_{LT}}{2A_r T_{0,r}^q(y)f_1}, \\
  & \langle \sin(2\varphi-\varphi_{LT}) \rangle_{U,LT} = -x\kappa_M k_{\perp M}^2\frac{ A_r \tilde T_{3,r}^q(y)g^\perp_{LT}}{2A_r T_{0,r}^q(y)f_1}, \\
  & \langle \cos(\varphi-2\varphi_{TT}) \rangle_{U,TT} =  x\kappa_M k_{\perp M} \frac{A_r T_{2,r}^q(y)f_{TT}}{A_r T_{0,r}^q(y)f_1}, \\
  & \langle \sin(\varphi-2\varphi_{TT}) \rangle_{U,TT} = -x\kappa_M k_{\perp M} \frac{A_r \tilde T_{3,r}^q(y)g_{TT}}{A_r T_{0,r}^q(y)f_1}, \\
  & \langle \cos(3\varphi-3\varphi_{TT}) \rangle_{U,TT} = x\kappa_M k_{\perp M}^3\frac{ A_r T_{2,r}^q(y)f^\perp_{TT}}{2A_r T_{0,r}^q(y)f_1}, \\
  & \langle \sin(3\varphi-2\varphi_{TT}) \rangle_{U,TT} = x\kappa_M k_{\perp M}^3\frac{A_r \tilde T_{3,r}^q(y)g^\perp_{TT}}{2A_r T_{0,r}^q(y)f_1}.
\end{align}
There are only ten azimuthal asymmetries left if the weak interactions are excluded.

\subsection{Azimuthal asymmetries form polarized electron beam}

For the case of the polarized electron beam, we obtain similar results as the unpolarized case.
They have one-to-one correspondence. At the leading twist we have six kinds of asymmetries,
\begin{align}
 & \langle \sin(\varphi-\varphi_S) \rangle_{L,T} = -\lambda_ek_{\perp M} \frac{ A_r \tilde T_{0,r}^q(y)f^\perp_{1T}}{2A_r T_{0,r}^q(y)f_1}, \\
 & \langle \cos(\varphi-\varphi_S) \rangle_{L,T} =  \lambda_ek_{\perp M} \frac{ A_r T_{1,r}^q(y)g^\perp_{1T}}{2A_r T_{0,r}^q(y)f_1}, \\
 & \langle \sin(\varphi-\varphi_{LT}) \rangle_{L,LT} = \lambda_e k_{\perp M} \frac{ A_r T_{1,r}^q(y)g^\perp_{1LT}}{2A_r T_{0,r}^q(y)f_1}, \\
 & \langle \cos(\varphi-\varphi_{LT}) \rangle_{L,LT} = \lambda_ek_{\perp M} \frac{A_r \tilde T_{0,r}^q(y)f^\perp_{1LT}}{2A_r T_{0,r}^q(y)f_1}, \\
 & \langle \sin(2\varphi-\varphi_{TT}) \rangle_{L,TT} = \lambda_ek_{\perp M}^2 \frac{A_r T_{1,r}^q(y)g^\perp_{1TT}}{2A_r T_{0,r}^q(y)f_1}, \\
 & \langle \cos(2\varphi-\varphi_{TT}) \rangle_{L,TT} = -\lambda_ek_{\perp M}^2 \frac{A_r \tilde T_{0,r}^q(y)f^\perp_{1TT}}{2A_r T_{0,r}^q(y)f_1}.
\end{align}
We find that three parity-violating azimuthal asymmetry modulations among the six in total are left if only electromagnetic interactions are considered.
At twist-3, we have 18 twist-3 azimuthal asymmetries. They are given by
\begin{align}
  & \langle \cos\varphi \rangle_{L,U} = \lambda_e x\kappa_M k_{\perp M} \frac{A_r \tilde T_{2,r}^q(y)f^\perp}{A_r T_{0,r}^q(y)f_1}, \\
  & \langle \sin\varphi \rangle_{L,U} = \lambda_e x\kappa_M k_{\perp M} \frac{ A_r T_{3,r}^q(y)g^\perp}{A_r T_{0,r}^q(y)f_1}, \\
  & \langle \cos\varphi \rangle_{L,L} = \lambda_e x\kappa_M k_{\perp M} \frac{ A_r \tilde T_{2,r}^q(y) f^\perp - \lambda_h A_r T_{3,r}^q(y)g^\perp_L}{A_r T_{0,r}^q(y)f_1}, \\
  & \langle \sin\varphi \rangle_{L,L} = \lambda_e x\kappa_M k_{\perp M} \frac{A_r T_{3,r}^q(y) g^\perp + A_r \tilde T_{2,r}^q(y)f^\perp_L}{A_r T_{0,r}^q(y)f_1}, \\
  & \langle \cos\varphi \rangle_{L,LL} = \lambda_e x\kappa_M k_{\perp M} \frac{A_r \tilde T_{2,r}^q(y)(f^\perp + S_{LL} f^\perp_{LL})}{A_r T_{0,r}^q(y)f_1}, \\
  & \langle \sin\varphi \rangle_{L,LL} = \lambda_e x\kappa_M k_{\perp M} \frac{A_r T_{3,r}^q(y)(g^\perp + S_{LL}g^\perp_{LL})}{A_r T_{0,r}^q(y)f_1}, \\
  & \langle \cos\varphi_S \rangle_{L,T} = - \lambda_e x\kappa_M\frac{ A_r T_{3,r}^q(y)g_T}{A_r T_{0,r}^q(y)f_1}, \\
  & \langle \sin\varphi_S \rangle_{L,T} =  \lambda_e x\kappa_M\frac{ A_r \tilde T_{2,r}^q(y)f_T}{A_r T_{0,r}^q(y)f_1}, \\
  & \langle \cos(2\varphi-\varphi_S) \rangle_{L,T} = -\lambda_e x\kappa_M k_{\perp M}^2\frac{ A_r T_{3,r}^q(y)g^\perp_T}{2A_r T_{0,r}^q(y)f_1}, \\
  & \langle \sin(2\varphi-\varphi_S) \rangle_{L,T} = \lambda_e x\kappa_M k_{\perp M}^2\frac{A_r \tilde T_{2,r}^q(y)f^\perp_T}{2A_r T_{0,r}^q(y)f_1}, \\
  & \langle \cos\varphi_{LT} \rangle_{L,LT} = \lambda_e x\kappa_M\frac{A_r \tilde T_{2,r}^q(y)f_{LT}}{A_r T_{0,r}^q(y)f_1}, \\
  & \langle \sin\varphi_{LT} \rangle_{L,LT} = \lambda_e x\kappa_M\frac{ A_r T_{3,r}^q(y)g_{LT}}{A_r T_{0,r}^q(y)f_1}, \\
  & \langle \cos(2\varphi-\varphi_{LT}) \rangle_{L,LT} = \lambda_e x\kappa_M k_{\perp M}^2\frac{ A_r \tilde T_{2,r}^q(y) f^\perp_{LT}}{2A_r T_{0,r}^q(y)f_1}, \\
  & \langle \sin(2\varphi-\varphi_{LT}) \rangle_{L,LT} = \lambda_ex\kappa_M k_{\perp M}^2\frac{ A_r T_{3,r}^q(y)g^\perp_{LT}}{2A_r T_{0,r}^q(y)f_1}, \\
  & \langle \cos(\varphi-2\varphi_{TT}) \rangle_{L,TT} = -\lambda_e x\kappa_M k_{\perp M}\frac{ A_r \tilde T_{2,r}^q(y)f_{TT}}{A_r T_{0,r}^q(y)f_1}, \\
  & \langle \sin(\varphi-2\varphi_{TT}) \rangle_{L,TT} = \lambda_e x\kappa_M k_{\perp M}\frac{ A_r T_{3,r}^q(y)g_{TT}}{A_r T_{0,r}^q(y)f_1}, \\
  & \langle \cos(3\varphi-3\varphi_{TT}) \rangle_{L,TT} = -\lambda_e x\kappa_M k_{\perp M}^3\frac{A_r \tilde T_{2,r}^q(y) f^\perp_{TT}}{2A_r T_{0,r}^q(y)f_1}, \\
  & \langle \sin(3\varphi-3\varphi_{TT}) \rangle_{L,TT} = -\lambda_e x\kappa_M k_{\perp M}^3\frac{A_r T_{3,r}^q(y)g^\perp_{TT}}{2A_r T_{0,r}^q(y)f_1}.
\end{align}
There are only ten azimuthal asymmetries left if the weak interactions are excluded.

\subsection{Parity-violating asymmetries}
With the advent of highly-polarized electron beams, parity violation measurements have become a standard tool for probing a variety of phenomena, for example, the Standard Model, the role of strange quarks in the proton and the neutron distribution in nuclei. The parity violating asymmetry in DIS offers a unique window into the interesting physics. This asymmetry is sensitive to the hadronic structure of the nucleon and to the Standard Model couplings, e.g., $c_A, c_3$. To be explicit, it is convenient to consider the inclusive DIS. Integrating over $d^2k_\perp'$ yields
\begin{widetext}
\begin{align}
  \frac{d\sigma_{in}^{ZZ}}{dx dy d\psi} =  \frac{\alpha_{\rm{em}}^2 \chi}{y Q^2}\Biggl\{& \left(T_0^q(y) -\lambda_e \tilde T_0^q(y) \right) \big[f_1(x)+S_{LL}f_{1LL}(x)\big] - \left(\tilde T_1^q(y) -\lambda_e T_1^q(y) \right) \lambda_h g_{1L}(x) \nonumber\\
  -|S_T|2x\kappa_M&\Big[\sin\varphi_S \left( T_2^q(y)-\lambda_e \tilde T_2^q(y)\right)f_T(x)- \cos\varphi_S \left( \tilde T_3^q(y)-\lambda_e T_3^q(y)\right)g_T(x) \Big]  \nonumber\\
  -|S_{LT}2x\kappa_M|&\Big[\sin\varphi_{LT} \left( \tilde T_3^q(y)-\lambda_e T_3^q(y)\right)g_{LT}(x) + \cos\varphi_{LT} \left( T_2^q(y)-\lambda_e \tilde T_2^q(y)\right)f_{LT}(x)\Big]
 \Biggr\}.
\end{align}
\end{widetext}
The complete differential cross section is given by
\begin{align}
\frac{d\sigma_{in}}{dx dy d\psi} = \frac{d\sigma_{in}^{ZZ} + d\sigma_{in}^{\gamma Z} + d\sigma_{in}^{\gamma\gamma}}{dx dy d\psi}.
\end{align}

To calculate the parity-violating asymmetries, we assume that the lepton is longitudinal polarized. Here we introduce the definition of the parity-violating asymmetry,
\begin{align}
  A_\sigma^{PV}=\frac{d\sigma_\sigma(\lambda_e=+1)-d\sigma_\sigma(\lambda_e=-1)}{d\sigma_U^{\gamma\gamma}}, \label{f:PVasymmetryDef}
\end{align}
where the subscript $\sigma$ denotes the target polarization, superscript $PV$ denotes parity-violating. First of all, we take the target as unpolarized. According to the definition, we have
\begin{align}
 & A_U^{PV}=-\frac{\big[\chi_{int}\tilde T_{0,\gamma Z}^q(y) +\chi\tilde T_{0}^q(y) \big]f_1(x)}{e_q^2 A(y) f_1(x)},
\end{align}
which is consistent with the calculation in Ref. \cite{Cahn:1977uu} at low energy.

It is also interesting to calculate the parity-violating asymmetries when the beam is unpolarized while the target is polarized. For the longitudinal polarized target, we have
\begin{align}
  & A_L^{PV}=-\frac{\big[\chi_{int}\tilde T_{1,\gamma Z}^q(y) +\chi\tilde T_{1}^q(y) \big]g_{1L}(x)}{e_q^2 A(y) f_1(x)}, \\
  & A_{LL}^{PV}=-\frac{\big[\chi_{int}\tilde T_{0,\gamma Z}^q(y) +\chi\tilde T_{0}^q(y) \big]f_{1LL}(x)}{e_q^2 A(y) f_1(x)}.
\end{align}

For the transverse polarized target, we have
\begin{align}
  & A_{T,x}^{PV}=2x\kappa_M\frac{\big[\chi_{int}\tilde T_{3,\gamma Z}^q(y) +\chi\tilde T_{3}^q(y) \big]g_{T}(x)}{e_q^2 A(y) f_1(x)}, \\
  & A_{T,y}^{PV}=-2x\kappa_M\frac{\big[\chi_{int}T_{2,\gamma Z}^q(y) +\chi T_{2}^q(y) \big]f_T(x)}{e_q^2 A(y) f_1(x)}, \\
  & A_{LT,x}^{PV}=-2x\kappa_M\frac{\big[\chi_{int} T_{2,\gamma Z}^q(y) +\chi T_{2}^q(y) \big]f_{LT}(x)}{e_q^2 A(y) f_1(x)}, \\
  & A_{LT,y}^{PV}=-2x\kappa_M\frac{\big[\chi_{int}\tilde T_{3,\gamma Z}^q(y) +\chi \tilde T_{3}^q(y) \big]g_{LT}(x)}{e_q^2 A(y) f_1(x)}.
\end{align}

Parity-violating asymmetries given in this part combine the electroweak and QCD theories. Measuring these asymmetries can be important ways to examine electroweak and QCD theories simultaneously.

\section{Summary} \label{sec:summary}

In this paper, we present a complete and systematic calculation of the parity-violating current jet production SIDIS process at the EIC.
We consider both the EM and weak interactions.
We presented the general form of the differential cross section of this process in terms of structure functions by making full kinematical analysis.
In QCD parton model the calculations are carried out by applying the collinear expansion where the multiple gluon scattering is taken into account and gauge links are obtained systematically and automatically.

We consider both unpolarized and polarized electron beams scattering off polarized spin-1 target.
There are in total 36 structure functions contribute at twist-2 and 72 structure functions contribute at twist-3 for different polarization configurations.
We also presented the azimuthal asymmetries results.
For both unpolarized and polarized electron beams cases, there are 24 azimuthal asymmetries up to twist-3,
in which 6 of them correspond to the leading twist TMD PDFs while the other 18 correspond to the twist-3 TMD PDFs.
Among these structure functions and azimuthal asymmetries results, only one fourth of them will left if only electromagnetic interaction is take into account.
The remaining others are all generated through weak interaction and its interference with EM interaction.
We also calculate the parity-violating asymmetries for weak interaction is considered. Though, the EIC is being proposed mainly for the study of strong interactions, it has a unique ability to measure parity violating quantities.

\section*{Acknowledgements}
We thank Professor Zuo-tang Liang for useful suggestions.
This work was supported by the National Natural Science Foundation of China (No. 11947055) and the National Laboratory Foundation (No. 6142004180203).


\begin{thebibliography}{0}

\bibitem{Collins:1989gx}
 See e.g., J.~C.~Collins, D.~E.~Soper and G.~F.~Sterman,
  Adv.\ Ser.\ Direct.\ High Energy Phys.\  {\bf 5}, 1 (1989)
  doi:10.1142/9789814503266$\_$0001
  [hep-ph/0409313].
  ¡°Perturbative Quantum Chromodynamics,¡± A.H. Mueller ed., Singapore, World Scientific, 1989.





\bibitem{Cahn:1977uu}
  R.~N.~Cahn and F.~J.~Gilman,
  Phys.\ Rev.\ D {\bf 17}, 1313 (1978).
  doi:10.1103/PhysRevD.17.1313.


\bibitem{Anselmino:1993tc}
  M.~Anselmino, P.~Gambino and J.~Kalinowski,
  Z.\ Phys.\ C {\bf 64}, 267 (1994)
  doi:10.1007/BF01557397
  [hep-ph/9401264].


\bibitem{Prescott:1978tm}
  C.~Y.~Prescott {\it et al.},
  Phys.\ Lett.\  {\bf 77B}, 347 (1978).
  doi:10.1016/0370-2693(78)90722-0.

\bibitem{Prescott:1979dh}
  C.~Y.~Prescott {\it et al.},
  Phys.\ Lett.\  {\bf 84B}, 524 (1979).
  doi:10.1016/0370-2693(79)91253-X.









\bibitem{Aniol:2004hp}
  K.~A.~Aniol {\it et al.} [HAPPEX Collaboration],
  Phys.\ Rev.\ C {\bf 69}, 065501 (2004)
  doi:10.1103/PhysRevC.69.065501
  [nucl-ex/0402004].

\bibitem{Aniol:2005zf}
  K.~A.~Aniol {\it et al.} [HAPPEX Collaboration],
  Phys.\ Rev.\ Lett.\  {\bf 96}, 022003 (2006)
  doi:10.1103/PhysRevLett.96.022003
  [nucl-ex/0506010].

\bibitem{Aniol:2005zg}
  K.~A.~Aniol {\it et al.} [HAPPEX Collaboration],
  Phys.\ Lett.\ B {\bf 635}, 275 (2006)
  doi:10.1016/j.physletb.2006.03.011
  [nucl-ex/0506011].


\bibitem{Armstrong:2005hs}
  D.~S.~Armstrong {\it et al.} [G0 Collaboration],
  Phys.\ Rev.\ Lett.\  {\bf 95}, 092001 (2005)
  doi:10.1103/PhysRevLett.95.092001
  [nucl-ex/0506021].

\bibitem{Androic:2009aa}
  D.~Androic {\it et al.} [G0 Collaboration],
  Phys.\ Rev.\ Lett.\  {\bf 104}, 012001 (2010)
  doi:10.1103/PhysRevLett.104.012001
  [arXiv:0909.5107 [nucl-ex]].


\bibitem{Wang:2013kkc}
  D.~Wang {\it et al.} [Jefferson Lab Hall A Collaboration],
  Phys.\ Rev.\ Lett.\  {\bf 111}, no. 8, 082501 (2013)
  doi:10.1103/PhysRevLett.111.082501
  [arXiv:1304.7741 [nucl-ex]].

\bibitem{Wang:2014guo}
  D.~Wang {\it et al.},
  Phys.\ Rev.\ C {\bf 91}, no. 4, 045506 (2015)
  doi:10.1103/PhysRevC.91.045506
  [arXiv:1411.3200 [nucl-ex]].

\bibitem{Anthony:2003ub}
  P.~L.~Anthony {\it et al.} [SLAC E158 Collaboration],
  Phys.\ Rev.\ Lett.\  {\bf 92}, 181602 (2004)
  doi:10.1103/PhysRevLett.92.181602
  [hep-ex/0312035].

\bibitem{Anthony:2005pm}
  P.~L.~Anthony {\it et al.} [SLAC E158 Collaboration],
  Phys.\ Rev.\ Lett.\  {\bf 95}, 081601 (2005)
  doi:10.1103/PhysRevLett.95.081601
  [hep-ex/0504049].

\bibitem{Spayde:1999qg}
  D.~T.~Spayde {\it et al.} [SAMPLE Collaboration],
  Phys.\ Rev.\ Lett.\  {\bf 84}, 1106 (2000)
  doi:10.1103/PhysRevLett.84.1106
  [nucl-ex/9909010].

\bibitem{Ito:2003mr}
  T.~M.~Ito {\it et al.} [SAMPLE Collaboration],
  Phys.\ Rev.\ Lett.\  {\bf 92}, 102003 (2004)
  doi:10.1103/PhysRevLett.92.102003
  [nucl-ex/0310001].

\bibitem{Maas:2004dh}
  F.~E.~Maas {\it et al.},
  Phys.\ Rev.\ Lett.\  {\bf 94}, 152001 (2005)
  doi:10.1103/PhysRevLett.94.152001
  [nucl-ex/0412030].


\bibitem{Maas:2004ta}
  F.~E.~Maas {\it et al.} [A4 Collaboration],
  Phys.\ Rev.\ Lett.\  {\bf 93}, 022002 (2004)
  doi:10.1103/PhysRevLett.93.022002
  [nucl-ex/0401019].

\bibitem{PVDIS:Jlab6}
  X.~Zheng, P.~Reimer, and E.~A.~R.~Michaels
  \url{http://www.jlab.org/exp_prog/proposals/08/PR-08-011.pdf}.

\bibitem{PVDIS:JLab12}
  P.~Reimer, X.~Zheng, and E.~A.~K.~Paschke \url{https://www.jlab.org/exp_prog/proposals/07/PR12-07-102.pdf}.

\bibitem{Accardi:2012qut}
  A.~Accardi {\it et al.},
  Eur.\ Phys.\ J.\ A {\bf 52}, no. 9, 268 (2016)
  doi:10.1140/epja/i2016-16268-9
  [arXiv:1212.1701 [nucl-ex]].


\bibitem{Ji:1993ey}
  X.~D.~Ji,
  Nucl.\ Phys.\ B {\bf 402}, 217 (1993).
  doi:10.1016/0550-3213(93)90642-3.


\bibitem{Anselmino:1994gn}
  M.~Anselmino, A.~Efremov and E.~Leader,
  Phys.\ Rept.\  {\bf 261}, 1 (1995)
  Erratum: [Phys.\ Rept.\  {\bf 281}, 399 (1997)]
  doi:10.1016/0370-1573(95)00011-5
  [hep-ph/9501369].

\bibitem{Boer:1999uu}
  D.~Boer, R.~Jakob and P.~J.~Mulders,
  Nucl.\ Phys.\ B {\bf 564}, 471 (2000)
  doi:10.1016/S0550-3213(99)00586-6
  [hep-ph/9907504].

\bibitem{Anselmino:2001ey}
  M.~Anselmino, M.~Boglione, U.~D'Alesio and F.~Murgia,
  Eur.\ Phys.\ J.\ C {\bf 21}, 501 (2001)
  doi:10.1007/s100520100741
  [hep-ph/0106055].

\bibitem{deFlorian:2012wk}
  D.~de Florian and Y.~Rotstein Habarnau,
  Eur.\ Phys.\ J.\ C {\bf 73}, no. 3, 2356 (2013)
  doi:10.1140/epjc/s10052-013-2356-3
  [arXiv:1210.7203 [hep-ph]].

\bibitem{Moreno:2014kia}
  O.~Moreno, T.~W.~Donnelly, J.~W.~Van Orden and W.~P.~Ford,
  Phys.\ Rev.\ D {\bf 90}, no. 1, 013014 (2014)
  doi:10.1103/PhysRevD.90.013014
  [arXiv:1406.4494 [hep-th]].


\bibitem{Mulders:1995dh}
  P.~J.~Mulders and R.~D.~Tangerman,
  Nucl.\ Phys.\ B {\bf 461}, 197 (1996)
  Erratum: [Nucl.\ Phys.\ B {\bf 484}, 538 (1997)]
  doi:10.1016/S0550-3213(96)00648-7, 10.1016/0550-3213(95)00632-X
  [hep-ph/9510301].


\bibitem{Bacchetta:2006tn}
  A.~Bacchetta, M.~Diehl, K.~Goeke, A.~Metz, P.~J.~Mulders and M.~Schlegel,
  JHEP {\bf 0702}, 093 (2007)
  doi:10.1088/1126-6708/2007/02/093
  [hep-ph/0611265].


\bibitem{Boer:1997mf}
  D.~Boer, R.~Jakob and P.~J.~Mulders,
  Nucl.\ Phys.\ B {\bf 504}, 345 (1997)
  doi:10.1016/S0550-3213(97)00456-2
  [hep-ph/9702281].


\bibitem{Ellis:1982wd}
  R.~K.~Ellis, W.~Furmanski and R.~Petronzio,
  Nucl.\ Phys.\ B {\bf 207}, 1 (1982).
  Nucl.\ Phys.\ B {\bf 212}, 29 (1983).


\bibitem{Qiu:1990xxa}
  J.~-w.~Qiu and G.~F.~Sterman,
  Nucl.\ Phys.\ B {\bf 353}, 105 (1991).

  Nucl.\ Phys.\ B {\bf 353}, 137 (1991).



\bibitem{Chen:2016moq}
  K.~b.~Chen, W.~h.~Yang, S.~y.~Wei and Z.~t.~Liang,
  Phys.\ Rev.\ D {\bf 94}, no. 3, 034003 (2016)
  doi:10.1103/PhysRevD.94.034003
  [arXiv:1605.07790 [hep-ph]].





\bibitem{Liang:2006wp}
  Z.~t.~Liang and X.~N.~Wang,
  Phys.\ Rev.\ D {\bf 75}, 094002 (2007)
  [hep-ph/0609225].

\bibitem{Song:2010pf}
 Y.~k.~Song, J.~h.~Gao, Z.~t.~Liang and X.~N.~Wang,
  Phys.\ Rev.\ D {\bf 83}, 054010 (2011)
  doi:10.1103/PhysRevD.83.054010
  [arXiv:1012.4179 [hep-ph]].


\bibitem{Wei:2013csa}
  S.~y.~Wei, Y.~k.~Song and Z.~t.~Liang,
  Phys.\ Rev.\ D {\bf 89}, no. 1, 014024 (2014)
  doi:10.1103/PhysRevD.89.014024
  [arXiv:1309.4191 [hep-ph]].

\bibitem{Wei:2014pma}
  S.~Y.~Wei, K.~b.~Chen, Y.~k.~Song and Z.~t.~Liang,
  Phys.\ Rev.\ D {\bf 91}, no. 3, 034015 (2015)
  doi:10.1103/PhysRevD.91.034015
  [arXiv:1410.4314 [hep-ph]].


\bibitem{Song:2013sja}
  Y.~k.~Song, J.~h.~Gao, Z.~t.~Liang and X.~N.~Wang,
  Phys.\ Rev.\ D {\bf 89}, no. 1, 014005 (2014)
  doi:10.1103/PhysRevD.89.014005
  [arXiv:1308.1159 [hep-ph]].

\bibitem{Wei:2016far}
  S.~y.~Wei, Y.~k.~Song, K.~b.~Chen and Z.~t.~Liang,
  Phys.\ Rev.\ D {\bf 95}, no. 7, 074017 (2017)
  doi:10.1103/PhysRevD.95.074017
  [arXiv:1611.08688 [hep-ph]].


\bibitem{Yang:2017sxz}
  W.~h.~Yang, K.~b.~Chen and Z.~t.~Liang,
  Phys.\ Rev.\ D {\bf 96}, no. 5, 054016 (2017)
  doi:10.1103/PhysRevD.96.054016
  [arXiv:1707.00402 [hep-ph]].

\bibitem{Yang:2020ezt}
  W.~Yang,
  Nucl.\ Phys.\ A {\bf 997}, 121729 (2020)
  doi:10.1016/j.nuclphysa.2020.121729.


\bibitem{Bacchetta:2000jk}
  A.~Bacchetta and P.~J.~Mulders,
  Phys.\ Rev.\ D {\bf 62}, 114004 (2000)
  doi:10.1103/PhysRevD.62.114004
  [hep-ph/0007120].


\bibitem{Diehl:2005pc}
  M.~Diehl and S.~Sapeta,
  Eur.\ Phys.\ J.\ C {\bf 41}, 515 (2005)
  doi:10.1140/epjc/s2005-02242-9
  [hep-ph/0503023].


\end{thebibliography}
\end{document}